\documentclass[aps, superscriptaddress, nofootinbib]{revtex4-2}

\usepackage{amsmath}
\usepackage{amssymb}
\usepackage{graphicx}

\usepackage{color}

\begin{document}

\title{Monodromy Approach to Pair Production of Charged Black Holes and Electric Fields}

\author{Chiang-Mei Chen} \email{cmchen@phy.ncu.edu.tw}
\affiliation{Department of Physics, National Central University, Chungli 32001, Taiwan}
\affiliation{Center for High Energy and High Field Physics (CHiP), National Central University, Chungli 32001, Taiwan}

\author{Toshimasa Ishige} \email{ishiget@yahoo.co.jp}
\affiliation{Graduate School of Science and Engineering, Chiba University, Chiba-shi 263-8522, Japan}

\author{Sang Pyo Kim} \email{sangkim@kunsan.ac.kr}
\affiliation{Department of Physics, Kunsan National University, Kunsan 54150, Korea}
\affiliation{Asia Pacific Center for Theoretical Physics (APCTP), Pohang 37673, Korea}

\author{Akitoshi Takayasu} \email{takitoshi@risk.tsukuba.ac.jp}
\affiliation{Faculty of Engineering, Information and Systems, University of Tsukuba, Ibaraki 305-8573, Japan}

\author{Chun-Yu Wei} \email{weijuneyu@gmail.com}
\affiliation{Department of Physics, National Central University, Chungli 32001, Taiwan}

\date{\today}

\begin{abstract}
To find the pair production, absorption cross section and quasi-normal modes in background fields, we advance the monodromy method that makes use of the regular singular points of wave equations. We find the mean number of pairs produced in background fields whose mode equations belong to the Riemann differential equation and apply the method to the three particular cases: (i) charges near the horizon of near-extremal black holes, (ii) charges with minimal energy under the static balance in nonextremal charged black holes, and (iii) charges in the Sauter-type electric fields. We then compare the results from the monodromy with those from the exact wave functions in terms of the hypergeometric functions with three regular singular points. The explicit elaboration of monodromy and the model calculations worked out here seem to reveal evidences that the monodromy may provide a practical technique to study the spontaneous pair production in general black holes and electromagnetic fields.
\end{abstract}


\maketitle

\section{Introduction}

One of nonperturbative aspects of quantum field theory is spontaneous particle production from background fields, two of whose most prominent phenomena are the (Sauter-)Schwinger mechanism in electromagnetic fields~\cite{Schwinger:1951nm, Sauter:1931zz} and the Hawking radiation in black holes~\cite{Hawking:1975vcx}. The physical concept behind the particle production is that the background fields change the vacuum in such a way that the out-vacuum is superposed of multiparticle states of the in-vacuum and vice versa~\cite{DeWitt:2003}. The Klein-Gordon equation for a scalar field, though a linear equation, has been solved only for a few background fields~\cite{Bagrov:1990}. Indeed it has been a challenge for a long time either to directly find the exact wave functions in terms of the special functions or to develop some approximation schemes for the wave functions, such as the WKB method~\cite{Bender:1999} or Borel-summed WKB method~\cite{Honda:2015}.

Particle production has been an interesting topic in cosmology, in particular, in expanding universes~\cite{Parker:1968mv} (for a recent review, see~\cite{Ford:2021syk}). Recently the Schwinger pair production of charged particles and antiparticles by a strong electric field has attracted much attention because ultra-strong lasers have been proposed to reach field strengths near the Schwinger field in the near future, and spontaneous production of electrons and positrons will be a direct test of QED in strong field region (for a review on astrophysical applications, see~\cite{Ruffini:2009hg} and for a recent review, see~\cite{Fedotov:2022ely}). Charged black holes are an arena in which both the Schwinger mechanism and the Hawking radiation intertwine to spontaneously emit charges.

The field equation for a charged scalar in charged black holes, such as the Reissner-Nordstr\"{o}m (RN) and Kerr-Newman (KN) black holes in an asymptotically flat or (anti-)de Sitter space (A)dS, has not been exactly solved yet in terms of special functions in the global covering space. A conventional wisdom has been to solve the field equation in the near-horizon region and the asymptotic region, and then to connect those wave functions. Another method is to use the enhanced symmetry of background geometry in some limits. The (near-)extremal black holes have a near-horizon geometry whose enhanced symmetry allows one to exactly solve wave functions. Two of us (CMC, SPK) have studied spontaneous production of charged particles from (near-)extremal RN or KN black holes in the asymptotically flat or (A)dS spaces~\cite{Chen:2012zn, Chen:2016caa, Chen:2017mnm, Chen:2020mqs, Zhang:2020apg, Cai:2020trh, Chen:2021jwy}.

To understand the emission of charges from charged black holes, one has to solve the Klein-Gordon or Dirac equation in the RN or KN black holes. However, the Klein-Gordon equation in nonextremal charged black holes is a confluent Heun equation, with three poles including a double pole at infinity, which in general cannot be analytically solved~\cite{Hsiao2019}. In this paper we will study an alternative technique, the so-called monodromy that directly computes the scattering coefficients from the information of poles without knowing the analytic solutions~\cite{IKSY2013, Castro:2013lba}. This technique can pass over the formidable task of finding the wave functions in terms of the special functions but give one useful information necessary for the mean number or absorption cross section of emitted particles from black holes. Moreover, it can also be used to calculate the quasi-normal modes by imposing a proper boundary condition.

As the first step in this direction, we find the mean number of pairs produced in background fields whose mode equations belong to the Riemann differential equation. We then consider the three special cases of the Riemann equation: (i) charges in near-extremal charged black holes, (ii) charges with minimal energy under a static balance in nonextremal black holes, and (iii) charges in electric fields that belong to a generalized P\"oschl-Teller potential. All the three cases have wave functions in terms of the hypergeometric functions with three regular singular points and with transformation formulas between different regions, and will provide one with models to test the monodromy approach. The phase-integral method that also makes use of the regular singular points~\cite{Kim:2007pm} gives the leading term for the pair production from near-extremal charged black holes~\cite{Chen:2012zn, Chen:2016caa, Chen:2017mnm} and explains the Stokes phenomenon~\cite{Kim:2013cka}. On the other hand, the monodromy method recovers the exact results for those models studied in this paper. Thus we argue that the monodromy seems a completion of the phase-integral method and may work for more general cases.

In section~\ref{sec:MonodromyBoundary}, we introduce the definition and properties of monodromy and impose the boundary conditions for pair production and absorption cross section. And we find the mean number of pair production in background fields whose mode equations are described by the Riemann differential function. In section~\ref{sec:SpecialCases}, we derive, case by case, the pair production from monodromy when the governing equations are the hypergeometric function with three regular singular points. One specific case is the emission of scalar charges from the near-horizon region of near-extremal RN black holes~\cite{Chen:2012zn}, and the other case is the scalar charges with minimal energy under a static balance in nonextremal RN black holes~\cite{Hsiao2019}, which is also given by the Riemann differential equation. In section~ \ref{sec:PoschlTeller}, we apply the monodromy to charges in electric fields of the generalized P\"{o}schl-Teller potential, one of which is the Sauter-type electric field. The monodromy confirms the pair production formula from the exact solutions of the wave equation. In section~\ref{sec:Conclusion}, we discuss the features and applications of the monodromy to more general cases.

\section{Monodromy and Boundary Conditions} \label{sec:MonodromyBoundary}

In this section, we are going to give a concise introduction to monodromy and to explain how to apply this technique to study pair production mechanisms. Such ideas had been discussed in~\cite{Castro:2013lba}, but the explanations here are more rigorous and complete, and, in particular, several obscure issues during the physical applications will be clarified. Thus the application will turn out to be just a straightforward process.


\subsection{Monodromy Representation of Fundamental Group}

Consider a second order ordinary differential equation (ODE) of the form
\begin{equation} \label{eq:ODE}
\frac{d^2}{dr^2} R(r) + U(r) \frac{d}{dr} R(r) + V(r) R(r) = 0,
\end{equation}
where $U(r)$ and $V(r)$ are meromorphic functions in $\mathbb{C}$ with finite poles. Let $r_i$ ($i = 1, 2, \dots, m$) be the whole set of poles of either $U(r)$ or $V(r)$. If both $(r - r_i) U(r)$ and $(r - r_i)^2 V(r)$ are holomorphic at $r_i$, then the singularity $r_i$ is called regular, otherwise irregular.

The ODE~\eqref{eq:ODE} can be extended to the Riemann sphere, i.e. $\mathbb{P} = \mathbb{C} \cup \{ \infty \}$. Setting $s = 1/r$ and $S(s) = R(r)$, the ODE~\eqref{eq:ODE} is transformed into
\begin{equation} \label{eq:ODEinfty}
\begin{split}
& \frac{d^2}{ds^2} S(s) + U_\infty(s) \frac{d}{ds} S(s) + V_\infty(s) S(s) = 0,
\\
U_\infty(s) &= 2 s^{-1} - s^{-2} U(1/s), \qquad V_\infty(s) = s^{-4} V(1/s).
\end{split}
\end{equation}
Note that $r_i \in \mathbb{C} \setminus \{ 0 \}$ is a pole of $U(r)$ or $V(r)$ if and only if $s_i = 1/r_i \in \mathbb{C} \setminus \{ 0 \}$ is a pole of $U_\infty(s)$ or $V_\infty(s)$. If $s = 0$ is a pole, then $r_{m + 1} = \infty$ is a singularity of~\eqref{eq:ODE} and the number of singularities in $\mathbb{P}$ becomes $n = m + 1$, otherwise $n = m$. If all of the singularities $r_1, \dots, r_n$ are regular, then the ODE~\eqref{eq:ODE} is called a \emph{Fuchsian differential equation}. Let $\mathbb{X} = \mathbb{P} \setminus \{ r_1, r_2, \dots, r_n \}$. It is clear that the solutions of the ODE~\eqref{eq:ODE} are holomorphic in $\mathbb{X}$, since they are differentiable everywhere in $\mathbb{X}$. However, since $\mathbb{X}$ is not simply connected, a solution of the ODE~\eqref{eq:ODE} may be a multivalued holomorphic function in $\mathbb{X}$. Such multivaluedness of the solutions is characterized by the \emph{monodromy} which reflects the topological property of $\mathbb{X}$.

\begin{figure}[htbp]
\centering
\includegraphics[scale=0.5]{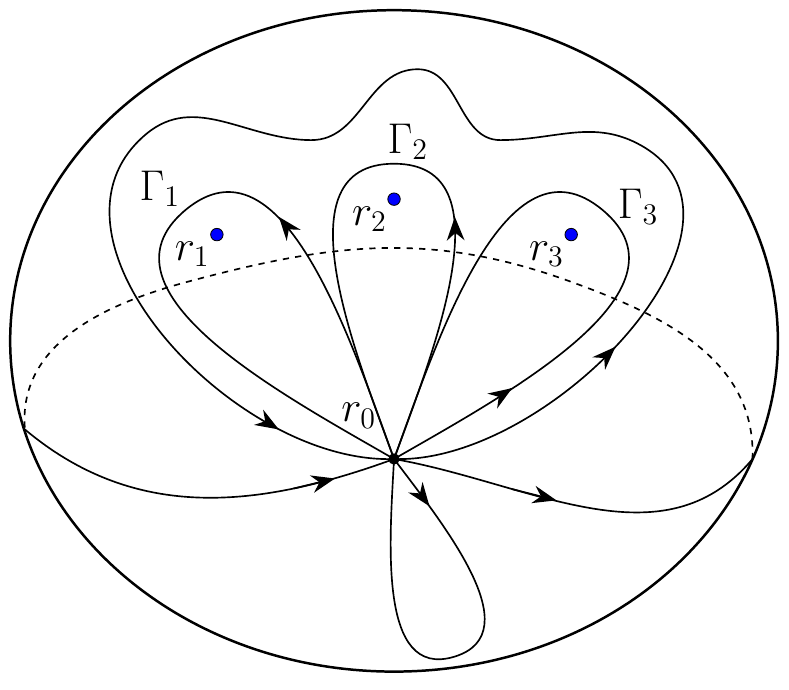}
\caption{A sketch of the loops on the Riemann sphere $\mathbb{P}$: The loops $\Gamma_1, \Gamma_2$, and $\Gamma_3$ from the base point $r_0$ are enclosing the singular points $r_1, r_2$, and $r_3$, respectively. $\Gamma_1 \cdot \Gamma_2 \cdot \Gamma_3$ continuously deforms to $r_0$ and yields the global relation~\eqref{global-relation-fundamental-group}.}
\label{fig:Deformation-of-the-loops}
\end{figure}

Fix a base point $r_0 \in \mathbb{X}$. Let $\Gamma_i$ ($i = 1, 2, \dots, n$) be a loop in $\mathbb{X}$ initiated at $r_0$ enclosing only $r_i$ counterclockwise. The product of loops $\Gamma_j \cdot \Gamma_i$ is defined to be a loop going firstly along $\Gamma_i$ then $\Gamma_j$. Moreover, $\pi_1(\mathbb{X}, r_0)$, the fundamental group of $\mathbb{X}$ with the base point $r_0$, is generated by $[\Gamma_1], \dots, [\Gamma_n]$ where $[\Gamma_i]$ denotes the homotopy class of loops for $r_0$ containing $\Gamma_i$.
Since the base point $r_0$ is a constant loop (i.e., the image of the constant map $[0, 1] \to \{ r_0 \}$), $[r_0]$ is the identity element of $\pi_1(\mathbb{X}, r_0)$ denoted by $1$. We can construct $\Gamma_1, \dots, \Gamma_n$ to be arranged clockwise in this order. Then, the composite loop $\Gamma_1 \cdot \Gamma_2 \cdots \Gamma_n$, which is homotopic to $r_0$ in $\mathbb{X}$ (i.e., it continuously deforms to $r_0$ in $\mathbb{X}$ as illustrated in Figure~\ref{fig:Deformation-of-the-loops} for the case of $n = 3$), gives the following \emph{global relation} of  $\pi_1(\mathbb{X}, r_0)$:
\begin{equation} \label{global-relation-fundamental-group}
[\Gamma_1] \cdot [\Gamma_2] \cdots [\Gamma_n] = 1.
\end{equation}

Let $\mathbb{D}$ be a simply connected open set in $\mathbb{X}$ with $r_0 \in \mathbb{D}$. Since $\mathbb{D}$ is simply connected, the \emph{monodromy theorem} in the complex analysis implies that all of the solutions of the ODE~\eqref{eq:ODE} are single-valued in $\mathbb{D}$. In virtue of the existence and uniqueness of solutions for the initial value problem, the map from the space of initial values at $r_0$, say $\left[ R(r_0), R^\prime(r_0) \right]^T = \mathbf{R}_0$ where $R^\prime = dR/dr$, to the solution space of the ODE~\eqref{eq:ODE} over $\mathbb{D}$
\begin{equation}
\Phi: \mathbb{C}^2 \to \mathrm{Sol}(\mathbb{D}), \qquad \mathbf{R}_0 \mapsto R(r),
\end{equation}
is a $\mathbb{C}$-linear isomorphism. Let $\mathbf{R}_0^{1}$ and $\mathbf{R}_0^{2}$ be $\mathbb{C}$-linearly independent vectors in $\mathbb{C}^2$, then $\Phi(\mathbf{R}_0^{1})$ and $\Phi(\mathbf{R}_0^{2})$ are $\mathbb{C}$-linearly independent solutions in $\mathrm{Sol}(\mathbb{D})$. Suppose that $\Gamma$ is the image of the smooth map
\begin{equation}
r_\Gamma: [0, 1] \to \mathbb{X}, \qquad t \mapsto r_\Gamma(t) \quad \text{with} \quad r_\Gamma(0) = r_\Gamma(1) = r_0.
\end{equation}
There exist $0 < t_1 < t_2 < 1$ such that $r_\Gamma\left( [0, t_1] \right) \subset \mathbb{D}$ and $r_\Gamma\left( [t_2, 1] \right) \subset \mathbb{D}$. As for $r_\Gamma([t_1,t_2])$, it is assumed to be any path in $\mathbb{X}$ connecting $r_\Gamma(t_1)$ and $r_\Gamma(t_2)$. In the course of the analytic continuation along $\Gamma$, the solutions $\Phi(\mathbf{R}_0^{1})$ and $\Phi(\mathbf{R}_0^{2})$ stay as they are while in $r_\Gamma\left( [0, t_1] \right)$,  and  after the analytic continuation along  $r_\Gamma\left([t_1,t_2]\right)$ when finally going into $r_\Gamma\left( [t_2, 1] \right)$, they consequently become respective $\mathbb{C}$-linear sums of $\Phi(\mathbf{R}_0^{1})$ and $\Phi(\mathbf{R}_0^{2})$.
Thus, the \emph{monodromy matrix} $\mathbf{M}_\Gamma$ along the loop $\Gamma$ with respect to $[\Phi(\mathbf{R}^1_0), \Phi(\mathbf{R}^2_0)]$ is defined by a nonsingular matrix such that
\begin{equation} \label{eq:monodormy-def}
\left[ \Phi(\mathbf{R}_0^1),\Phi( \mathbf{R}_0^2) \right] \, \mathbf{M}_\Gamma = \Gamma_\ast \left[\Phi( \mathbf{R}^1_0), \Phi(\mathbf{R}^2_0) \right],
\end{equation}
where $\Gamma_\ast$ denotes the analytic continuation along $\Gamma$.

Let us transform~\eqref{eq:ODE} into the \emph{Pfaffian form}
\begin{equation} \label{eq:pfaffian-form}
\frac{d}{dr} \begin{bmatrix} R \\ R^\prime \end{bmatrix} = \begin{bmatrix} 0 & 1 \\ -V & -U \end{bmatrix} \begin{bmatrix} R \\ R^\prime \end{bmatrix} \quad \iff \quad \frac{d}{dr} \mathbf{R} = \mathbf{A} \mathbf{R}.
\end{equation}
Using the discretization of~\eqref{eq:pfaffian-form}, the \emph{path-ordered-exponential} of $\mathbf{A}$ along $\Gamma$ is defined by
\begin{equation}\label{eq:path-ordered-exponential}
\begin{split}
\mathbf{L}_\Gamma &= \mathcal{P} \exp\left( \int_\Gamma \mathbf{A}(r) dr \right) = \mathcal{P} \exp\left( \int_0^1 \mathbf{A}(r_\Gamma(t)) r_\Gamma^\prime(t) dt \right)
\\
&= \lim_{N \to \infty} \left( \mathbf{I} + \mathbf{A}\left( r_\Gamma(t_{N-1}) \right) r_\Gamma^\prime(t_{N-1}) \Delta t \right) \cdots
\left( \mathbf{I} + \mathbf{A}\left( r_\Gamma(t_0) \right) r_\Gamma^\prime(t_0) \Delta t \right),
\end{split}
\end{equation}
where $\Delta t = 1/N$, $t_k = k/N$. It is clear that
\begin{equation}\label{eq:homomorphism-of-POE}
\mathbf{L}_{\Gamma^\prime \cdot \Gamma} = \mathbf{L}_{\Gamma^\prime} \mathbf{L}_\Gamma
\end{equation}
holds. Applying $\Phi^{-1}$ to the both sides of \eqref{eq:monodormy-def}, we have
\begin{equation} \label{eq:monodromy-computation}
\left[ \mathbf{R}_0^1, \mathbf{R}_0^2 \right] \mathbf{M}_\Gamma = \mathbf{L}_\Gamma \left[ \mathbf{R}_0^1, \mathbf{R}_0^2 \right]
\quad \iff \quad
\mathbf{M}_\Gamma = \left[ \mathbf{R}_0^1, \mathbf{R}_0^2 \right]^{-1} \mathbf{L}_\Gamma \left[ \mathbf{R}_0^1, \mathbf{R}_0^2 \right].
\end{equation}
Thus, the monodromy matrix is similar to the path-ordered-exponential, which is denoted by $\mathbf{M}_\Gamma \sim \mathbf{L}_\Gamma$.

The \emph{monodromy representation} of the fundamental group is given by
\begin{equation}
\rho: \pi_1(\mathbb{X}, r_0) \to \mathrm{GL}(2, \mathbb{C}), \qquad [\Gamma] \mapsto \mathbf{M}_\Gamma.
\end{equation}
From~\eqref{eq:homomorphism-of-POE} and~\eqref{eq:monodromy-computation}, it holds that
\begin{equation}
\rho([\Gamma^\prime] \cdot [\Gamma]) = \mathbf{M}_{\Gamma^\prime \cdot \Gamma} = \mathbf{M}_{\Gamma^\prime} \mathbf{M}_\Gamma = \rho([\Gamma^\prime]) \rho([\Gamma]).
\end{equation}
Therefore, $\rho$ is a homomorphism of groups. Here the image of $\rho$ is called the \emph{monodromy group}.
From the global relation of the fundamental group~\eqref{global-relation-fundamental-group}, we obtain the global relation of the monodromy group such that
\begin{equation}
\mathbf{M}_{\Gamma_1} \mathbf{M}_{\Gamma_2} \cdots \mathbf{M}_{\Gamma_n} = \rho([\Gamma_1]) \rho([\Gamma_2]) \cdots \rho([\Gamma_n]) = \rho([\Gamma_1] \cdot [\Gamma_2] \cdots [\Gamma_n]) = \rho(1) = \mathbf{I}.
\end{equation}
In particular, $\mathbf{L}_{\Gamma_1} \mathbf{L}_{\Gamma_2} \cdots \mathbf{L}_{\Gamma_n} = \mathbf{I}$ holds with respect to $[\mathbf{R}_0^1, \mathbf{R}_0^2] = \mathbf{I}$.

Set $\hat{U}_i(r) = (r - r_i) U(r)$ and $\hat{V_i}(r) = (r - r_i)^2 V(r)$ for $i = 1, 2, \dots, m$. If $r_{m + 1} = \infty$ is a singularity,
set $\hat{U}_\infty(r) = 2 - r U(r)$ and $\hat{V}_\infty(r) = r^{2} V(r)$.
Then the ODE~\eqref{eq:ODE} is transformed as follows;
\begin{equation}\label{eq:transform}
\begin{split}
\text{ODE~\eqref{eq:ODE}} &\quad \iff \quad \frac{d^2R}{dr^2} + \frac{\hat{U}_i}{(r - r_i)} \frac{dR}{dr} + \frac{\hat{V}_i}{(r - r_i)^2} R = 0
\\
&\quad \iff \quad
(r - r_i)^2 \frac{d^2R}{dr^2} + \hat{U}_i (r - r_i) \frac{dR}{dr} + \hat{V}_i R = 0.
\end{split}
\end{equation}
We introduce the Euler operator $\delta_i = (r - r_i) d/dr$. Using $(r - r_i)^2 d^2/dr^2 = \delta_i^2 - \delta_i$, as a sequel to the transformation in~\eqref{eq:transform} we obtain
\begin{equation}\label{eq:ODE-Euler}
\delta_i^2 R + (\hat{U}_i - 1) \delta_i R + \hat{V} R = 0.
\end{equation}
Thus, we have another Pfaffian form:
\begin{equation}\label{eq:pfaffian-form-Euler}
\delta_i \begin{bmatrix} R \\ \delta_i R \end{bmatrix} = \begin{bmatrix} 0 & 1 \\ -\hat{V}_i & 1 - \hat{U}_i \end{bmatrix} \begin{bmatrix} R \\ \delta_i R \end{bmatrix} \quad \iff \quad \delta_i \hat{\mathbf{R}} = \hat{\mathbf{A}}_i \hat{\mathbf{R}}.
\end{equation}
Using the discretization of~\eqref{eq:pfaffian-form-Euler}, the path-ordered-exponential of $\hat{\mathbf{A}}_i$ along $\Gamma$ is defined by
\begin{equation}\label{eq:path-ordered-exponantilal-Euler}
\begin{split}
(\hat{\mathbf{L}}_i)_\Gamma &=
\mathcal{P} \exp\left( \int_\Gamma \hat{\mathbf{A}}_i(r) d \log(r - r_i) \right)
= \mathcal{P} \exp\left( \int_0^1 \hat{\mathbf{A}}_i(r_\Gamma(t)) \frac{r_\Gamma^\prime(t)}{r_\Gamma(t) - r_i} dt \right)
\\
&=
\lim_{N \to \infty} \left( \mathbf{I} + \hat{\mathbf{A}}_i\left( r_\Gamma(t_{N - 1}) \right) \frac{r_\Gamma^\prime(t_{N-1}) \Delta t}{r_\Gamma(t_{N-1}) - r_i} \right) \cdots
\left( \mathbf{I} + \hat{\mathbf{A}}_i\left( r_\Gamma(t_0) \right) \frac{r_\Gamma^\prime(t_0) \Delta t}{r_\Gamma(t_0)-r_i} \right),
\end{split}
\end{equation}
where $\Delta t = 1/N,$ and $t_k = k/N$.

In virtue of the existence and uniqueness of the solutions for the initial value problem with the initial value $(R, \delta_i R)^T = \hat{\mathbf{R}}_0$ for~\eqref{eq:ODE-Euler}, the map
\begin{equation}
\hat{\Phi}_i: \mathbb{C}^2 \to \mathrm{Sol}(\mathbb{D}), \qquad \hat{\mathbf{R}}_0 \mapsto R(r),
\end{equation}
is a $\mathbb{C}$-linear isomorphism.
Here we suppose $[\Phi(\mathbf{R}_0^1), \Phi(\mathbf{R}_0^2)] = [\hat{\Phi}_i(\hat{\mathbf{R}}_0^1), \hat{\Phi}_i(\hat{\mathbf{R}}_0^2)]$.
Then, applying $\hat{\Phi}_i^{-1}$ on the both sides of~\eqref{eq:monodormy-def}, we have
\begin{equation} \label{eq:monodromy-computation-Euler}
\left[ \hat{\mathbf{R}}_0^1, \hat{\mathbf{R}}_0^2 \right] \mathbf{M}_\Gamma = (\hat{\mathbf{L}}_i)_\Gamma \left[ \hat{\mathbf{R}}_0^1, \hat{\mathbf{R}}_0^2 \right]
\quad \iff \quad
\mathbf{M}_\Gamma = \left[ \hat{\mathbf{R}}_0^1, \hat{\mathbf{R}}_0^2 \right]^{-1} (\hat{\mathbf{L}}_i)_\Gamma \left[ \hat{\mathbf{R}}_0^1, \hat{\mathbf{R}}_0^2 \right].
\end{equation}
For a sufficiently small number $\epsilon > 0$, one may let $\Gamma_i$ continuously deform to $\Lambda_{0i}(\epsilon)^{-1} \cdot \Lambda_i(\epsilon) \cdot \Lambda_{0i}(\epsilon)$ where $\Lambda_{0i}(\epsilon)$ is a path connecting $r_0$ to $r_i + \epsilon$, and $\Lambda_i(\epsilon)$ is a circle initiated at $r_i + \epsilon$ encircling $r_i$ counterclockwise. Thus
\begin{equation}\label{eq:lasso-representation}
(\hat{\mathbf{L}}_i)_{\Gamma_i} = (\hat{\mathbf{L}}_i)_{\Lambda_{0i}(\epsilon)}^{-1} (\hat{\mathbf{L}}_i)_{\Lambda_i(\epsilon)} (\hat{\mathbf{L}}_i)_{\Lambda_{0i}(\epsilon)}.
\end{equation}
If $r_i$ is a regular singularity, the entries of $\hat{\mathbf{A}}_i(r_i)$ are finite and it follows from~\eqref{eq:path-ordered-exponantilal-Euler} that
\begin{equation} \label{eq:infenitesimal-circle}
\lim_{\epsilon \to 0} \left( \hat{\mathbf{L}}_i \right)_{\Lambda_i(\epsilon)} = \lim_{N \to \infty} \left( \mathbf{I} + \frac{2 \pi i \hat{\mathbf{A}}_i(r_i)}{N} \right)^N = \mathrm{e}^{2 \pi i \hat{\mathbf{A}}_i(r_i)}.
\end{equation}
In summary, we obtain from~\eqref{eq:monodromy-computation},~\eqref{eq:monodromy-computation-Euler},~\eqref{eq:lasso-representation} and~\eqref{eq:infenitesimal-circle} that
\begin{equation}
\mathbf{M}_{\Gamma_i} \sim \mathbf{L}_{\Gamma_i} \sim \left( \hat{\mathbf{L}}_i \right)_{\Gamma_i} \sim \mathrm{e}^{2\pi i \hat{\mathbf{A}}_i(r_i)}.
\end{equation}

Suppose the ODE~\eqref{eq:ODE} is a Fuchsian differential equation and the number of singularities is three, i.e., $n = 3$. Then it is known that if $\mathbf{M}_1, \mathbf{M}_2, \mathbf{M}_3 \in \mathrm{GL}(2, \mathbb{C})$ satisfy
\begin{equation} \label{eq:MMMI}
\mathbf{M}_1 \sim \mathrm{e}^{2 \pi i \hat{\mathbf{A}}_1(r_1)}, \quad \mathbf{M}_2 \sim \mathrm{e}^{2 \pi i \hat{\mathbf{A}}_2(r_2)}, \quad \mathbf{M}_3 \sim \mathrm{e}^{2 \pi i \hat{\mathbf{A}}_3(r_3)}, \quad \text{and} \quad \mathbf{M}_1 \mathbf{M}_2 \mathbf{M}_3 = \mathbf{I},
\end{equation}
then $\mathbf{M}_1 = \mathbf{M}_{\Gamma_1}, \mathbf{M}_2 = \mathbf{M}_{\Gamma_2}$ and $\mathbf{M}_3 = \mathbf{M}_{\Gamma_3}$ for some initial value $[\mathbf{R}_0^1, \mathbf{R}_0^2]$. It is worth noting that this is no longer true for $n > 3$.
The characteristic equation of $\hat{\mathbf{A}}_i(r_i)$:
\begin{equation}\label{eq:indicial-equation}
\det\left( x \mathbf{I} - \hat{\mathbf{A}}_i(r_i) \right) = x^2 - \mathrm{tr} \hat{\mathbf{A}}_i(r_i) x + \det\hat{\mathbf{A}}_i(r_i) = x^2 - \left( 1 - \hat{U}_i(r_i) \right) x + \hat{V}_i(r_i) = 0,
\end{equation}
and its roots, i.e., the eigenvalues are respectively called the \emph{indicial equation} and the \emph{characteristic exponents} of $r_i$ for the ODE~\eqref{eq:ODE}.
Let $\mathbf{\rho} = (\rho^1, \rho^2)^T$, $\mathbf{\sigma} = (\sigma^1, \sigma^2)^T$ and $\mathbf{\tau} = (\tau^1, \tau^2)^T$  be the characteristic exponents of $r_1 , r_2$ and $r_3$ respectively. It is known that they satisfy the \emph{Fuchsian relation}
\begin{equation} \label{eq:Fuchsian-relation}
\rho^1 + \rho^2 + \sigma^1 + \sigma^2 + \tau^1 + \tau^2 = n - 2 = 1.
\end{equation}
Conversely, the characteristic exponents $\rho, \sigma, \tau \in \mathbb{C}^2$ of the singularities $r_1, r_2, r_3$ satisfying the Fuchsian relation~\eqref{eq:Fuchsian-relation} exactly determine the ODE~\eqref{eq:ODE}: if $r_3 = \infty$,
\begin{equation} \label{eq:Riemann-equation}
\begin{split}
\frac{d^2 R}{dr^2} &+ \left(\frac{1 - \rho^1 - \rho^2}{r - r_1} + \frac{1 - \sigma^1 - \sigma^2}{r - r_2}\right) \frac{dR}{dr}
\\
&+ \frac{1}{(r - r_1)(r - r_2)} \left( \frac{\rho^1 \rho^2 (r_1 - r_2)}{r - r_1} + \frac{\sigma^1 \sigma^2 (r_2 - r_1)}{r - r_2} + \tau^1 \tau^2 \right) R = 0,
\end{split}
\end{equation}
is called the \emph{Riemann differential equation}, whose solution is the Riemann P-function
\begin{equation} \label{eq:Riemann-scheme}
P \begin{pmatrix}
r_1 & r_2 & \infty & \\
\rho^1 & \sigma^1 & \tau^1 &; r \\
\rho^2 & \sigma^2 & \tau^2 &
\end{pmatrix}.
\end{equation}
The monodromy of  the Riemann differential equation~\eqref{eq:Riemann-equation} is given in Theorem 4.3.2 of~\cite{IKSY2013}. Namely, if $\rho^i + \sigma^j + \tau^k \not \in \mathbb{Z} \quad (i, j, k = 1, 2)$, then the monodromy representation $\rho$ is expressed up to conjugacy by the following matrices:
\begin{equation} \label{eq:rhoGamma}
\rho([\Gamma_1]) =
\begin{bmatrix} \mathrm{e}^{2 \pi i \rho^1} & 1 \\ 0 & \mathrm{e}^{2 \pi i \rho^2} \end{bmatrix}, \quad
\rho([\Gamma_2]) =
\begin{bmatrix} \mathrm{e}^{2 \pi i \sigma^1} & 0 \\ b & \mathrm{e}^{2 \pi i \sigma^2} \end{bmatrix}, \quad
\rho([\Gamma_3]) = \left( \rho([\Gamma_1]) \rho([\Gamma_2]) \right)^{-1},
\end{equation}
where
\begin{equation}
b = \mathrm{e}^{-2 \pi i \tau^1} + \mathrm{e}^{-2 \pi i \tau^2} - \mathrm{e}^{2 \pi i (\rho^1 + \sigma^1)} - \mathrm{e}^{2 \pi i (\rho^2 + \sigma^2)}.
\end{equation}
The monodromy group is generated by $\rho([\Gamma_1])$ and $\rho([\Gamma_2])$. In particular, $\rho([\Gamma_3])$ is determined by the global relation $\rho([\Gamma_1]) \rho([\Gamma_2]) \rho([\Gamma_3]) = \mathrm{I}$. Since $\rho$ is up to conjugacy, $\mathbf{C}^{-1} \rho(\cdot) \mathbf{C}$ is another monodromy representation of \eqref{eq:Riemann-equation} for any $\mathbf{C} \in \mathrm{GL}(2, \mathbb{C})$. If $[\mathbf{R}_0^1, \mathbf{R}_0^2]$ is the initial value at $r_0$ for $\rho$, then $[\mathbf{R}_0^1, \mathbf{R}_0^2] \mathbf{C}$ is the initial value at $r_0$ for $\mathbf{C}^{-1} \rho(\cdot) \mathbf{C}$. In general, the conjugacy class of the monodromy representation is uniquely determined by the ODE~\eqref{eq:ODE}, and called the \emph{monodromy} of the ODE~\eqref{eq:ODE}.

\subsection{Connection Matrix of Pair Production and Quasi-normal Modes}

Let $[\mathbf{F}_i^1, \mathbf{F}_i^2]$ be a pair of eigenvectors for eigenvalues $\alpha_i^1$ and $\alpha_i^2$ of $\mathbf{L}_{\Gamma_i}$.
We assume $\alpha_i^1 \neq \alpha_i^2$. Then $\Phi(\mathbf{F}_i^1)$ and $\Phi(\mathbf{F}_i^2)$ span $\mathrm{Sol}(\mathbb{D})$. It follows from~\eqref{eq:monodromy-computation} that
\begin{equation}
\begin{split}
& \mathbf{L}_{\Gamma_i} [\mathbf{F}_i^1, \mathbf{F}_i^2] = [\mathbf{F}_i^1, \mathbf{F}_i^2]
\begin{bmatrix}
\alpha_i^1 &  \\
 & \alpha_i^2 \\
\end{bmatrix}
\\
\iff \quad &
[\mathbf{R}_0^1, \mathbf{R}_0^2] \mathbf{M}_{\Gamma_i} [\mathbf{R}_0^1, \mathbf{R}_0^2]^{-1} [\mathbf{F}_i^1, \mathbf{F}_i^2] = [\mathbf{F}_i^1, \mathbf{F}_i^2]
\begin{bmatrix}
\alpha_i^1 &  \\
 & \alpha_i^2 \\
\end{bmatrix}
\\
\iff \quad &
\mathbf{M}_{\Gamma_i} [\mathbf{R}_0^1, \mathbf{R}_0^2]^{-1} [\mathbf{F}_i^1, \mathbf{F}_i^2] = [\mathbf{R}_0^1, \mathbf{R}_0^2]^{-1} [\mathbf{F}_i^1, \mathbf{F}_i^2]
\begin{bmatrix}
\alpha_i^1 &  \\
 & \alpha_i^2 \\
\end{bmatrix}.
\end{split}
\end{equation}
Thus it holds that $[\mathbf{E}_i^1, \mathbf{E}_i^2] = [\mathbf{R}_0^1, \mathbf{R}_0^2]^{-1} [\mathbf{F}_i^1, \mathbf{F}_i^2]$ is a pair of eigenvectors for eigenvalues $\alpha_i^1$ and $\alpha_i^2$ of $\mathbf{M}_{\Gamma_i}$ as well.

Furthermore, the \emph{connection matrix} $\mathbf{P}_i^j$ from $[\Phi(\mathbf{F}_i^1), \Phi(\mathbf{F}_i^2)]$ to
$[\Phi(\mathbf{F}_j^1), \Phi(\mathbf{F}_j^2)]$ is defined by
\begin{equation} \label{eq:ConnectionMatrix}
\begin{split}
& \left[ \Phi(\mathbf{F}_j^1), \Phi(\mathbf{F}_j^2) \right] = \left[ \Phi(\mathbf{F}_i^1), \Phi(\mathbf{F}_i^2) \right] \mathbf{P}_i^j
\\
\iff \quad &
[\mathbf{F}_j^1, \mathbf{F}_j^2] = [\mathbf{F}_i^1, \mathbf{F}_i^2] \, \mathbf{P}_i^j
\\
\iff \quad &
\mathbf{P}_i^j = [\mathbf{F}_i^1, \mathbf{F}_i^2]^{-1} [\mathbf{F}_j^1, \mathbf{F}_j^2] = [\mathbf{E}_i^1, \mathbf{E}_i^2]^{-1} [\mathbf{E}_j^1, \mathbf{E}_j^2].
\end{split}
\end{equation}
For any $\mathbf{P} = \left[\begin{smallmatrix} p_{11} & p_{12} \\ p_{21} & p_{22} \end{smallmatrix} \right] \in \mathrm{GL}(2, \mathbb{C})$, let us define
\begin{equation}
\begin{split}
& \mu(\mathbf{P}) \stackrel{\mathrm{def}}= |p_{11} p_{22}| - |p_{12} p_{21}|,    \\
& \nu(\mathbf{P}) \stackrel{\mathrm{def}}= p_{11} p_{22} p_{12}^\ast p_{21}^\ast - p_{11}^\ast p_{22}^\ast p_{12} p_{21}.
\end{split}
\end{equation}
If and only if $\mu(\mathbf{P}_i^j) > 0$ and $\nu(\mathbf{P}_i^j) = 0$, one can normalize $\mathbf{P}_i^j$ to be in $\mathrm{SU}(1, 1) = \{ \mathbf{P} = \left[ \begin{smallmatrix} p_{11} & p_{12} \\ p_{21} & p_{22} \end{smallmatrix} \right] \in \mathrm{GL}(2, \mathbb{C}) \mid p_{11} = p_{22}^\ast, p_{12} = p_{21}^\ast, \det \mathbf{P} = 1\}$  by rescaling the eigenvectors with factors $d_1, d_2, d_3, d_4 \in \mathbb{C}$, i.e.,
\begin{equation} \label{eq:normalized-ConnectionMatrix}
\hat{\mathbf{P}}_i^j =
\begin{bmatrix}
d_1 & \\
 & d_2 \\
\end{bmatrix}
\mathbf{P}_i^j
\begin{bmatrix}
d_3 & \\
 & d_4 \\
\end{bmatrix}
 \in \mathrm{SU}(1,1).
\end{equation}
If $\mu(\mathbf{P}_i^j) < 0$, then we can redefine $\mathbf{P}_i^j = [\mathbf{F}_i^2, \mathbf{F}_i^1]^{-1} [\mathbf{F}_j^1, \mathbf{F}_j^2]$ or  $[\mathbf{F}_i^1, \mathbf{F}_i^2]^{-1} [\mathbf{F}_j^2, \mathbf{F}_j^1]$ so that $\mu(\mathbf{P}_i^j) > 0$, since
\begin{equation}
\begin{split}
\mu\left( [\mathbf{F}_i^1, \mathbf{F}_i^2]^{-1} [\mathbf{F}_j^1, \mathbf{F}_j^2] \right) &= \mu\left( [\mathbf{F}_i^2, \mathbf{F}_i^1]^{-1} [\mathbf{F}_j^2, \mathbf{F}_j^1] \right) = - \mu\left( [\mathbf{F}_i^2, \mathbf{F}_i^1]^{-1} [\mathbf{F}_j^1, \mathbf{F}_j^2] \right)
\\
&= - \mu\left( [\mathbf{F}_i^1, \mathbf{F}_i^2]^{-1} [\mathbf{F}_j^2, \mathbf{F}_j^1] \right).
\end{split}
\end{equation}
As for the condition $\nu(P_i^j) = 0$, we will examine below under the physical assumptions.
After the normalization, the following ambiguity of phases $\theta_1, \theta_2, \theta_3, \theta_4 \in \mathbb{R}$ with $\theta_1 + \theta_2 + \theta_3 + \theta_4 = 0$ remains;
\begin{equation}
\hat{\mathbf{P}}_i^j  \in \mathrm{SU}(1,1) \quad \iff \quad
\begin{bmatrix}
\mathrm{e}^{i \theta_1} & \\
 & \mathrm{e}^{i \theta_2} \\
\end{bmatrix}
\hat{\mathbf{P}}_i^j
\begin{bmatrix}
\mathrm{e}^{i \theta_3} & \\
 & \mathrm{e}^{i \theta_4} \\
\end{bmatrix}
\in \mathrm{SU}(1,1).
\end{equation}

\begin{figure}[htbp]
\centering
\includegraphics[scale=0.2]{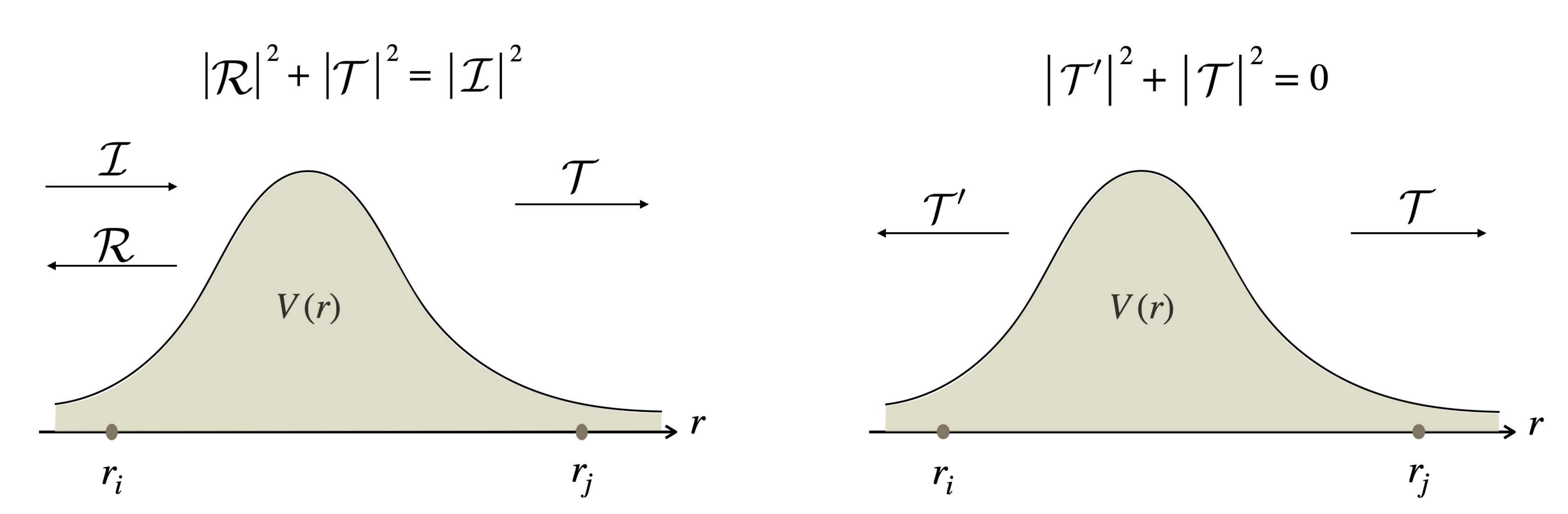}
\caption{The boundary condition for pair production [left] and quasi-normal mode [right] that are created in the region $r_i < r < r_j$ and to be observed at $r_j$. The in-going mode $\phi_i$ with amplitude ${\cal I}$, the out-going mode $\phi_i^*$ with amplitude ${\cal R}$ at $r_i$, and the out-going mode $\phi_j$ with amplitude ${\cal T}$ describe all particles. The complex conjugate of these modes describes the in-going mode $\phi_i$ with ${\cal R}^*$ and the out-going mode $\phi_i^*$ with ${\cal I}^*$ for particles at $r_i$, all forward in time, and the in-going mode $\phi_j^*$ with ${\cal T}^*$ for a particle at $r_j$ backward in time, that is, an antiparticle forward in time \cite{Hansen:1980nc, Kim:2000un} (for review, see Ref.~\cite{Brout:1995rd}). Hence the flux ratio $|\mathcal{T}/\mathcal{R}|^2$ gives the mean number for pair production.}  \label{fig:BoundaryCondition}
\end{figure}

Now, we are going to impose physical boundary conditions associated with pair production.
For the pair production created in the region $r_i < r < r_j$ which is to be observed at $r_j$ (see the left panel of Fig.~\ref{fig:BoundaryCondition}), one imposes the vanishing in-going (left-moving) modes at $r_j$. The ``incident'' flux (right-moving modes at $r_i$), $\mathcal{I}$, corresponds to virtual pairs of quantum fluctuation, the ``reflected'' flux (left-moving modes at $r_i$), $\mathcal{R}$, represents the reannihilation of virtual pairs and then the ``transmitted'' flux (right-moving modes at $r_j$), $\mathcal{T}$, manifests the particle production. The fluxes are conserved, namely $|\mathcal{R}|^2 + |\mathcal{T}|^2 = |\mathcal{I}|^2$.

We will assume that the ODE~\eqref{eq:ODE} is a Fuchsian differential equation, and for physical applications, all the coefficients in $U(r)$ and $V(r)$ and the singularities $r_1 < \dots < r_{n-1}$ are real, and $r_{n-1} > 0, r_n = \infty$. We will further assume that the discriminant of the indicial equation for $r_i (i = 1, 2, \dots, n)$ is negative, i.e., $\mathrm{tr}\hat{\mathbf{A}}_i(r_i)^2 - 4 \det \hat{\mathbf{A}}_i(r_i) < 0$. Then the characteristic exponents are $s_i^\pm = \beta_i \pm i \alpha_i$ with $\beta_i, \alpha_i \in \mathbb{R}$ and $\alpha_i \ne 0$.
Consider the Frobenius solutions $F_i^\pm = c_i^\pm (r - r_i)^{s_i^\pm} (\sum_{k \ge 0} c_{i, k}^{\pm} (r - r_i)^k)$ at the singularity $r_i  \ne \infty$, and $F_n^\pm = c_n^\pm r^{-s_n^\pm} (\sum_{k \ge 0} c_{n, k}^{\pm} r^{-k})$ at $r_n = \infty$.
From the hypothesis that all the coefficients of $U(r)$ and $V(r)$ are real, we may assume all $c_{i, k}$ are real.
It is known that for $r_i \ne \infty$ the radius of convergence for $\sum_{k \ge 0} c_{i,k}^{\pm} (r - r_i)^k$ is $R_i = \min_{k \ne i}  |r_k - r_i|$, and therefore the disc of convergence is $\mathbb{D}_i = \{ r \in \mathbb{C} \mid |r - r_i| < R_i \}$.
As for $r_n = \infty$, the radius of convergence for $\sum_{k \ge 0} c_{n, k}^{\pm} r^{-k}$ is $R_n = \max_{k \ne n} |r_k|$, and the disc of convergence is $\mathbb{D}_n = \{ r \in \mathbb{C} \mid |r - r_i| > R_n \}$.
For adjacent singularities $r_i, r_j$ with $j = i + 1$, we fix the base point $r_0 $ such that $r_i < r_0 < r_j$. We set $c_i^\pm = 1$, and $c_j^\pm = 1$ if $j = n$ and $c_j^\pm = (-1)^{-s_i^\pm}$ if $j < n$. Then if $r_0 \in \mathbb{D}_i \cap \mathbb{D}_j$ we immediately have $F_i^{+} = (F_i^{-})^*$ and $F_j^{+} = (F_j^{-})^*$ at $r = r_0$, otherwise analytically continue $F_i^\pm$ and $F_j^\pm$ to $r_0$ along the real axis using~\eqref{eq:path-ordered-exponential} with the integrands being real. Consequently, $F_i^\pm$ and $F_j^\pm$ have the respective initial values $[(\mathbf{F}_i^{+})_0, (\mathbf{F}_i^{-})_0]$ with $(\mathbf{F}_i^{+})_0 = (\mathbf{F}_i^{-})_0^\ast$ and $[(\mathbf{F}_j^{+})_0, (\mathbf{F}_j^{-})_0]$ with $(\mathbf{F}_j^{+})_0 = (\mathbf{F}_j^{-})_0^\ast$ at $r_0$, which are the respective pairs of eingenvectors of $\mathbf{L}_{\Gamma_i}$ and $\mathbf{L}_{\Gamma_j}$. Thus, the connection matrix is obtained by
\begin{equation}
\mathbf{P}_i^j = [(\mathbf{F}_i^{+})_0, (\mathbf{F}_i^{-})_0]^{-1} [(\mathbf{F}_j^{+})_0, (\mathbf{F}_j^{-})_0].
\end{equation}
If $\mu(\mathbf{P}_i^j) < 0$, we redefine $\mathbf{P}_i^j$ so that $\mu(\mathbf{P}_i^j) > 0$ as mentioned above.
On the other hand,
\begin{equation}
(\mathbf{P}_i^j)^\ast = [(\mathbf{F}_i^{-})_0, (\mathbf{F}_i^{+})_0]^{-1} [(\mathbf{F}_j^{-})_0, (\mathbf{F}_j^{+})_0]
= \begin{bmatrix} 0 & 1 \\ 1 & 0 \end{bmatrix} \mathbf{P}_i^j \begin{bmatrix} 0 & 1 \\ 1 & 0 \end{bmatrix}
\quad \Longrightarrow \quad \nu(\mathbf{P}_i^j) = 0.
\end{equation}
Thus $\mathbf{P}_i^j$ can be normalized to be in $\mathrm{SU}(1,1)$.

Therefore, the connection matrix $\mathbf{P}_i^j$ for pair production can be expressed as
\begin{equation} \label{eq:BoundaryCondition}
\mathbf{P}_i^j = \begin{bmatrix} \mathcal{I}/\mathcal{T} & \mathcal{R}/\mathcal{T} \\ \mathcal{R}^*/\mathcal{T}^* & \mathcal{I}^*/\mathcal{T}^* \end{bmatrix},
\end{equation}
which had been derived in~\cite{Castro:2013lba}. The conservation of flux implies the determinant of the connection matrix $\mathbf{P}_+^\infty$, up to an irrelevant phase, should be one.
Fortunately, the choice of the permutation of the eigenvectors when forcing $\mu(\mathbf{P}_i^j) > 0$ and the ambiguity of phases in~\eqref{eq:normalized-ConnectionMatrix} become irrelevant when computing $|\mathcal{I}|, \, |\mathcal{R}|$ and $|\mathcal{T}|$.

Additionally, we can also derive the connection matrix for the quasi-normal mode (QNM) with the boundary condition of only ingoing flux $\mathcal{T}'$ at $r_i$ (generally horizon) and outgoing flux $\mathcal{T}$ at $r_j$ (infinity) (see the right panel of Fig.~\ref{fig:BoundaryCondition}). In others words, we only need to replace $\mathcal{R}$ with $\mathcal{T}'$ and take $\mathcal{I} = 0$. Hence, the connection matrix of QNM will be
\begin{equation} \label{eq:QNM}
\mathbf{P}_i^j = \begin{bmatrix} 0 & \mathcal{T'}/\mathcal{T} \\ \mathcal{T'}^*/\mathcal{T}^* & 0 \end{bmatrix},
\end{equation}
which has vanishing diagonal components and $\det(\mathbf{P}_i^j) = 1$. A similar approach had been discussed in~\cite{Castro:2013lba}.

In the following sections we will apply this method to compute the pair production in a few physical backgrounds.

\subsection{Algorithm of Monodromy Approach}
Here we summarize the algorithm for applying monodromy method to compute the pair production and quasi-normal modes in background fields whose wave functions are given by the Riemann differential equation:
\begin{itemize}
\item determine the characteristic exponents:~\eqref{eq:Riemann-equation}
\item construct monodromy matrices:~\eqref{eq:rhoGamma}
\item compute connection matrix from eigenvectors of monodromy matrices:~\eqref{eq:ConnectionMatrix}
\item normalize connection matrix:~\eqref{eq:normalized-ConnectionMatrix}
\item read out physical quantities: pair production~\eqref{eq:BoundaryCondition} and QNM~\eqref{eq:QNM}.
\end{itemize}

\subsection{Mean Number of Pair Production in the Riemann Differential Equation}
We now consider physical problems whose mode equations are described by the Riemann differential equation~\eqref{eq:Riemann-equation} with the Riemann scheme~\eqref{eq:Riemann-scheme} and give a general formula for pair production. To be specific, let us consider pair production from $r = r_2$ to $r = r_\infty$, in which the imaginary part of $\sigma^1$ and $\tau^1$ (similarly the sign of $\sigma^2$ and $\tau^2$) should have the same sign. The associated monodromy matrices are
\begin{eqnarray}
&& \mathbf{M}_1 = \begin{bmatrix} \mathrm{e}^{2 \pi i \rho^1} & 1 \\ 0 & \mathrm{e}^{2 \pi i \rho^2} \end{bmatrix}, \qquad \mathbf{M}_2 = \begin{bmatrix} \mathrm{e}^{2 \pi i \sigma^1} & 0 \\ b & \mathrm{e}^{2 \pi i \sigma^2} \end{bmatrix},
\nonumber\\
&& \mathbf{M}_\infty = \begin{bmatrix} \mathrm{e}^{- 2 \pi i (\rho^1 + \sigma^1)} & - \mathrm{e}^{- 2 \pi i (\rho^1 + \sigma^1 + \rho^2)} \\ - b \, \mathrm{e}^{- 2 \pi i (\rho^1 + \sigma^1 + \sigma^2)} & \mathrm{e}^{- 2 \pi i (\rho^2 + \sigma^2)} + b \, \mathrm{e}^{- 2 \pi i (\rho^1 + \sigma^1 + \rho^2 + \sigma^2)} \end{bmatrix},
\end{eqnarray}
where
\begin{equation}
b = \mathrm{e}^{-2 \pi i \tau^1} + \mathrm{e}^{-2 \pi i \tau^2} - \mathrm{e}^{2 \pi i (\rho^1 + \sigma^1)} - \mathrm{e}^{2 \pi i (\rho^2 + \sigma^2)}.
\end{equation}
The eigenvalues of $\mathbf{M}_2$ and $\mathbf{M}_\infty$ are $(\mathrm{e}^{2 \pi i \sigma^1}, \mathrm{e}^{2 \pi i \sigma^2})$ and $(\mathrm{e}^{2 \pi i \tau^2}, \mathrm{e}^{2 \pi i \tau^1})$, respectively, and the corresponding eigenvectors are
\begin{equation}
\mathbf{E}_2 = \begin{bmatrix} \mathrm{e}^{2 \pi i \sigma^1} - \mathrm{e}^{2 \pi i \sigma^2} & 0 \\ b & 1 \end{bmatrix}, \qquad \mathbf{E}_\infty = \begin{bmatrix} \mathrm{e}^{2 \pi i \sigma^2} & \mathrm{e}^{2 \pi i \sigma^2} \\ \mathrm{e}^{2 \pi i (\rho^2 + \sigma^2)} - \mathrm{e}^{- 2 \pi i \tau^1} & \mathrm{e}^{2 \pi i (\rho^2 + \sigma^2)} - \mathrm{e}^{- 2 \pi i \tau^2} \end{bmatrix}.
\end{equation}
Hence, the connection matrix between $r = r_2$ to $r = \infty$ is
\begin{eqnarray}
\mathbf{P}_2^\infty &=& \begin{bmatrix} d_1 & 0 \\ 0 & d_2 \end{bmatrix} (\mathbf{E}_2)^{-1} \mathbf{E}_\infty \begin{bmatrix} d_3 & 0 \\ 0 & d_4 \end{bmatrix} = \frac1{\mathrm{e}^{2 \pi i \sigma^1} - \mathrm{e}^{2 \pi i \sigma^2}} \begin{bmatrix} d_1 d_3 \, \mathrm{e}^{2 \pi i \sigma^2} & d_1 d_4 \, \mathrm{e}^{2 \pi i \sigma^2} \\ - d_2 d_3 \, \Xi_1 & - d_2 d_4 \, \Xi_2 \end{bmatrix},
\\
&& \Xi_1 = \mathrm{e}^{2 \pi i (\sigma^1 - \tau^1)} + \mathrm{e}^{2 \pi i (\sigma^2 - \tau^2)}  - \mathrm{e}^{2 \pi i (\rho^1 + \sigma^1 + \sigma^2)} - \mathrm{e}^{2 \pi i (\rho^2 + \sigma^2 + \sigma^1)},
\nonumber\\
&& \Xi_2 = \mathrm{e}^{2 \pi i (\sigma^1 - \tau^2)} + \mathrm{e}^{2 \pi i (\sigma^2 - \tau^1)} - \mathrm{e}^{2 \pi i (\rho^1 + \sigma^1 + \sigma^2)} - \mathrm{e}^{2 \pi i (\rho^2 + \sigma^2 + \sigma^1)}.
\end{eqnarray}
To normalize the connection matrix $\mathbf{P}_2^\infty$ to the unit determinant, one should impose the constraint
\begin{equation}
d_1 d_2 d_3 d_4 = - \frac{\left( \mathrm{e}^{2 \pi i \sigma^1} - \mathrm{e}^{2 \pi i \sigma^2} \right)^2}{ \mathrm{e}^{2 \pi i \sigma^2} \left[ \mathrm{e}^{2 \pi i (\sigma^1 - \tau^2)} - \mathrm{e}^{2 \pi i (\sigma^1 - \tau^1)} + \mathrm{e}^{2 \pi i (\sigma^2 - \tau^1)} - \mathrm{e}^{2 \pi i (\sigma^2 - \tau^2)} \right]},
\end{equation}
which has the following relation to the fluxes for the pair production boundary condition
\begin{equation}
\mathbf{P}_2^\infty = \begin{bmatrix} \mathcal{I}/\mathcal{T} & \mathcal{R}/\mathcal{T} \\ \mathcal{R}^*/\mathcal{T}^* & \mathcal{I}^*/\mathcal{T}^* \end{bmatrix}.
\end{equation}
Finally, the mean number of pair production via the ``tunneling''  process is given by
\begin{equation} \label{eq_N}
{\cal N} = \frac{|\mathcal{T}|^2}{|\mathcal{R}|^2} = - \frac{\left( \mathrm{e}^{2 \pi i \sigma^1} - \mathrm{e}^{2 \pi i \sigma^2} \right)^2}{d_1 d_2 d_3 d_4 \, \mathrm{e}^{2 \pi i \sigma^2} \, \Xi_1} = \frac{\mathrm{e}^{2 \pi i (\sigma^1 - \tau^2)} - \mathrm{e}^{2 \pi i (\sigma^1 - \tau^1)} + \mathrm{e}^{2 \pi i (\sigma^2 - \tau^1)} - \mathrm{e}^{2 \pi i (\sigma^2 - \tau^2)}}{\mathrm{e}^{2 \pi i (\sigma^1 - \tau^1)} + \mathrm{e}^{2 \pi i (\sigma^2 - \tau^2)}  - \mathrm{e}^{2 \pi i (\rho^1 + \sigma^1 + \sigma^2)} - \mathrm{e}^{2 \pi i (\rho^2 + \sigma^2 + \sigma^1)}}.
\end{equation}
The mean number can be written as
\begin{equation} \label{eq_Nformula}
{\cal N} = \frac{\cos \pi (\sigma^1- \sigma^2 + \tau^1 - \tau^2) - \cos \pi (\sigma^1- \sigma^2 - \tau^1 + \tau^2)}{\cos \pi (\sigma^1- \sigma^2 - \tau^1 + \tau^2) + \cos \pi (\rho^1 - \rho^2)}.
\end{equation}

\section{Pair Production from Charged Black Holes: Two Special Cases} \label{sec:SpecialCases}

To investigate the Hawking radiation and the Schwinger effect, we consider a massive charged scalar field $\Phi$ carrying charge $q$ and mass $m$ which satisfies the Klein-Gordon (KG) equation
\begin{equation} \label{eq:KGeq}
\left( D_\nu D^\nu - m^2 \right) \Phi = 0,
\end{equation}
where the covariant derivative is defined as $D_\mu = \nabla_\mu - i q A_\mu$. This field propagates in the background of the RN black holes with charge $Q$ and mass $M$
\begin{equation} \label{eq:RNBH}
ds^2 = - f(r) dt^2 + \frac{dr^2}{f(r)} + r^2 d\Omega_2^2, \qquad A_\mu dx^\mu = - \frac{Q}{r} dt,
\end{equation}
where $f(r) = (r - r_+) (r - r_-)/r^2$ with $r_\pm = M \pm \sqrt{M^2 - Q^2}$. By separating the frequency and harmonics
\begin{equation}
\Phi(t, r, \theta, \phi) = \int d\omega \sum_{l n} \mathrm{e}^{ i n \phi} S_{l n}(\theta) R_{l n}(\omega, r) \mathrm{e}^{- i \omega t} ,
\end{equation}
where $\mathrm{e}^{ i n \phi} S_{l n}(\theta)$ are the spherical harmonics, the KG equation~(\ref{eq:KGeq}) reduces to an ODE
\begin{equation} \label{eq:radialKGeq}
f \frac{d}{d r} \left( r^2 f \frac{d}{d r} R \right) + \left[ (\omega r - q Q)^2 - (m^2 r^2 + \lambda_l) f \right] R = 0,
\end{equation}
where $R = R_{l n}(\omega, r)$ and $\lambda_l$ is the separation constant $\lambda_l = l (l + 1)$ for an integer $l$ ($l = 0, 1, \cdots$).

The Eq.~(\ref{eq:radialKGeq}) is a confluent Heun equation with three poles including a degenerate pole at spatial infinity. There are some technical difficulties in applying the monodromy approach. Here we consider two special cases such that the associated Klein-Gordon equation reduces to the Riemann differential equations which can be directly analyzed by the monodromy technique.

\subsection{Near Horizon of Near Extremal Black Holes} \label{subsec:NearExtremal}

\begin{figure}[htbp]
\centering
\includegraphics[scale=0.3]{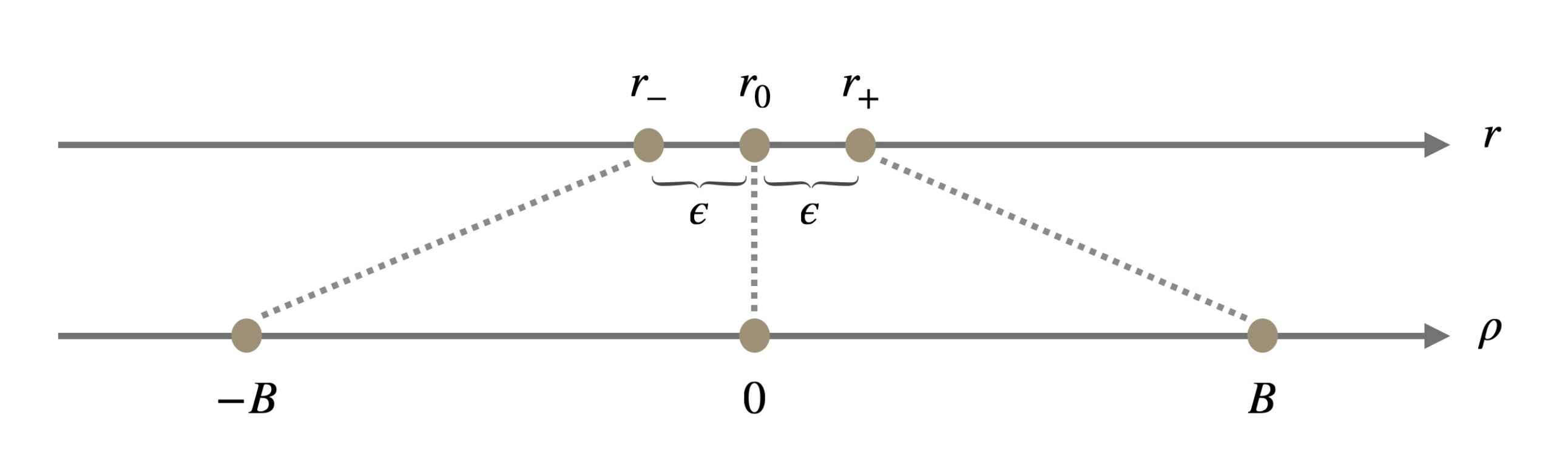}
\caption{Re-scaling of the coordinates near the horizon of near-extremal charged black holes. \label{fig:RescaleCoordinate}}
\end{figure}

Let's firstly focus on pair production in the near-horizon region of near-extremal RN black holes~\cite{Chen:2012zn}. By performing the following coordinate re-scaling, as shown in Fig.~\ref{fig:RescaleCoordinate}, and taking $\epsilon \to 0$
\begin{equation}
r \rightarrow Q + \epsilon \rho, \qquad M \rightarrow Q + \frac{\epsilon^2 B^2}{2 Q}, \qquad  t \rightarrow \frac{\tau}{\epsilon},
\end{equation}
the metric and the electromagnetic potential take the form in the $\mathrm{AdS}_2 \times S^2$ space
\begin{equation}
ds^2 = - \frac{\rho^2 - B^2}{Q^2} d\tau^2 + \frac{Q^2}{\rho^2 - B^2} d\rho^2 + Q^2 d\Omega_2^2, \qquad A_\mu dx^\mu = - \frac{\rho}{Q} d\tau,
\end{equation}
which $B$ is the black holes horizon and $-B$ the inner horizon in the new scaled coordinates. Hence, the radial part of the KG equation in the near-horizon region
\begin{equation} \label{eq:KGeqNearHorizon}
\frac{d^2 R}{d\rho^2} + \frac{2 \rho}{\rho^2 - B^2} \frac{dR}{d\rho} + \left[ \frac{Q^2 (\omega Q - q \rho)^2}{(\rho^2 - B^2)^2} - \frac{m^2 Q^2 + \lambda_l}{\rho^2 - B^2} \right] R = 0,
\end{equation}
is  described by a Riemann differential equation~\eqref{eq:Riemann-equation} with three regular singular points at $\rho_\pm = \pm B$ and infinity $\rho_\infty = \infty$.

To construct the monodromy matrices $\mathbf{M}_i$ we should firstly compute the characteristic exponents $s_\rightarrow^i$ and $s_\leftarrow^i$ at every singular point, $\rho_i, i = \pm, \infty$. By the indicial equation~ \eqref{eq:indicial-equation}, we can straightforwardly figure out the exponents and find the solution in the Riemann P-function
\begin{equation}
P \begin{pmatrix} \rho_- & \rho_+ & \infty & \\ s_\rightarrow^- & s_\rightarrow^+ & s_\rightarrow^\infty & ; \rho \\ s_\leftarrow^- & s_\leftarrow^+ & s_\leftarrow^\infty & \end{pmatrix} = P \begin{pmatrix} -B & B & \infty & \\ i \alpha_- & i \alpha_+ & \frac12 + i \alpha_\infty & ; \rho \\ - i \alpha_- & - i \alpha_+ & \frac12 - i \alpha_\infty & \end{pmatrix},
\end{equation}
where
\begin{equation} \label{eq:alphas}
2 \alpha_\pm = Q^2 \frac{\omega}{B} \mp q Q, \qquad \alpha_\infty = \sqrt{(q^2 - m^2) Q^2 - \lambda_l - 1/4}.
\end{equation}
Moreover, the property~(\ref{eq:MMMI}) requires that each monodromy matrix $\mathbf{M}_i$ should have two invariants
\begin{equation}
\det \mathbf{M}_i = \mathrm{e}^{i 2 \pi \left( s_\rightarrow^i + s_\leftarrow^i \right)} = 1, \quad \text{for} \quad i = \pm, \, \infty,
\end{equation}
and
\begin{equation}
\mathrm{tr} \, \mathbf{M}_i = \mathrm{e}^{i 2 \pi s_\rightarrow^i} + \mathrm{e}^{i 2 \pi s_\leftarrow^i} = \begin{cases} \quad 2 \cosh(2 \pi \alpha_i) & \mathrm{for} \; i = \pm, \\ -2 \cosh(2 \pi \alpha_\infty) & \mathrm{for} \; i = \infty. \end{cases}
\end{equation}
According to~\eqref{eq:rhoGamma} and~\eqref{eq:MMMI}, we assume the monodromy matrices as follows:
\begin{equation}
\begin{split}
& \mathbf{M}_- = \begin{bmatrix} \mathrm{e}^{-2 \pi \alpha_-} & 1 \\ 0 & \mathrm{e}^{2 \pi \alpha_-} \end{bmatrix}, \qquad \mathbf{M}_+ = \begin{bmatrix} \mathrm{e}^{-2 \pi \alpha_+} & 0 \\ b & \mathrm{e}^{2 \pi \alpha_+} \end{bmatrix},
\\
&\mathbf{M}_\infty = \begin{bmatrix} \mathrm{e}^{2 \pi (\alpha_+ + \alpha_-)} & - \mathrm{e}^{2 \pi \alpha_+} \\ -b \, \mathrm{e}^{2 \pi \alpha_-} & b + \mathrm{e}^{-2 \pi (\alpha_+ + \alpha_-)} \end{bmatrix},
\end{split}
\end{equation}
with
\begin{equation}
b = - \mathrm{e}^{2 \pi \alpha_\infty} - \mathrm{e}^{-2 \pi \alpha_\infty} - \mathrm{e}^{2 \pi (\alpha_+ + \alpha_-)} - \mathrm{e}^{-2 \pi (\alpha_+ + \alpha_-)}.
\end{equation}
Then, it is easy to figure out the eigenvectors with eigenvalues $\left\{ \mathrm{e}^{2 \pi \alpha_+}, \mathrm{e}^{-2 \pi \alpha_+} \right\}$ for the monodromy matrix $\mathbf{M}_+$
\begin{equation}
\mathbf{E}_+ = \begin{bmatrix} \mathbf{E}_+^1, \mathbf{E}_+^2 \end{bmatrix} = \begin{bmatrix} 0 & \mathrm{e}^{-2 \pi \alpha_+} - \mathrm{e}^{2 \pi \alpha_+} \\ 1 & b \end{bmatrix},
\end{equation}
and those with $\left\{ - \mathrm{e}^{-2 \pi \alpha_\infty}, - \mathrm{e}^{2 \pi \alpha_\infty} \right\}$ for the monodromy matrix $\mathbf{M}_\infty$
\begin{equation}
\mathbf{E}_\infty = \begin{bmatrix} \mathrm{e}^{2 \pi \alpha_+} & \mathrm{e}^{2 \pi \alpha_+} \\ \mathrm{e}^{-2 \pi \alpha_\infty} + \mathrm{e}^{2 \pi (\alpha_+ + \alpha_-)} & \mathrm{e}^{2 \pi \alpha_\infty} + \mathrm{e}^{2 \pi (\alpha_+ + \alpha_-)} \end{bmatrix}.
\end{equation}
Following~\eqref{eq:ConnectionMatrix} and~\eqref{eq:normalized-ConnectionMatrix}, it is straightforward to compute the connection matrix $\mathbf{P}_+^\infty$
\begin{equation}
\begin{split}
\mathbf{P}_+^\infty &= \begin{bmatrix} d_1 & 0 \\ 0 & d_2 \end{bmatrix} (\mathbf{E}_+)^{-1} \mathbf{E}_\infty \begin{bmatrix} d_3 & 0 \\ 0 & d_4 \end{bmatrix}
\\
&= \begin{bmatrix} - d_1 d_3 \frac{\cosh(2 \pi \alpha_-) + \cosh(2 \pi \alpha_+ + 2 \pi \alpha_\infty)}{\sinh(2 \pi\alpha_+)} & - d_1 d_4 \frac{\cosh(2 \pi \alpha_-) + \cosh(2 \pi \alpha_+ - 2 \pi \alpha_\infty)}{\sinh(2 \pi\alpha_+)} \\ - d_2 d_3 \frac{2 \mathrm{e}^{2 \pi \alpha_+}}{\sinh(2 \pi\alpha_+)} & - d_2 d_4 \frac{2 \mathrm{e}^{2 \pi \alpha_+}}{\sinh(2 \pi\alpha_+)} \end{bmatrix}.
\end{split}
\end{equation}
To normalize the connection matrix $\mathbf{P}_+^\infty$ to the unit determinant, one should impose the constraint
\begin{equation}
d_1 d_2 d_3 d_4 = \frac{\sinh^2(2 \pi \alpha_+)}{4 \mathrm{e}^{2 \pi \alpha_+} \sinh(2 \pi \alpha_+) \sinh(2 \pi \alpha_\infty)}.
\end{equation}
Furthermore, from Eq.~(\ref{eq:BoundaryCondition}), the absorption cross section ratio (the inverse of product of diagonal elements) is independent of $d_i$ as shown
\begin{equation} \label{eq:Thyper}
\sigma_\mathrm{Near} = \frac{\left| \mathcal{T}_\mathrm{Near} \right|^2}{\left| \mathcal{I}_\mathrm{Near} \right|^2} = \frac{\sinh(2 \pi \alpha_+) \sinh(2 \pi \alpha_\infty)}{\cosh(\pi \alpha_+ + \pi \alpha_- + \pi \alpha_\infty) \cosh(\pi \alpha_+ - \pi \alpha_- + \pi \alpha_\infty)},
\end{equation}
and, similarly, the mean number is
\begin{equation} \label{eq:Nhyper}
\mathcal{N}_\mathrm{Near} = \frac{\left| \mathcal{T}_\mathrm{Near} \right|^2}{\left| \mathcal{R}_\mathrm{Near} \right|^2} = \frac{\sinh(2 \pi \alpha_+) \sinh(2 \pi \alpha_\infty)}{\cosh(\pi \alpha_+ + \pi \alpha_- - \pi \alpha_\infty) \cosh(\pi \alpha_+ - \pi \alpha_- - \pi \alpha_\infty)},
\end{equation}
for the pair production from the near-horizon region of near-extremal black holes. Indeed, the mean number can be obtained directly from the general formula~\eqref{eq_Nformula} with $\rho^1 = i \alpha_-, \rho^2 = - i \alpha_-, \sigma^1 = i \alpha_+, \sigma^2 = - i \alpha_+$ and $\tau^1 = 1/2 + i \alpha_\infty, \tau^2 = 1/2 - i \alpha_\infty$. These results are identical with those obtained in~\cite{Chen:2012zn} by a different approach that used the transformation formulas of the Whittaker and hypergeometric functions.

To compute the QNM connection matrix~\eqref{eq:QNM}, we should firstly use $\det(\mathbf{P}_i^j) = 1$ to determine $d_1 d_2 d_3 d_4$ and then the property of vanishing diagonal components gives the condition, which is equivalent to $|\mathcal{I}|^2 \to 0$,
\begin{equation}
\frac{\cosh(\pi\alpha_- + \pi \alpha_+ + \pi \alpha_\infty) \cosh(\pi \alpha_- - \pi \alpha_+ - \pi \alpha_\infty)}{\sinh(2 \pi \alpha_+) \sinh(2 \pi \alpha_\infty)} = 0,
\end{equation}
where $\alpha_\pm$ in $\omega$~\eqref{eq:alphas} are function of frequency. Therefore, the condition for the QNM frequencies is
\begin{equation}
\cosh(\pi\alpha_- + \pi \alpha_+ + \pi \alpha_\infty) = 0 \quad \Rightarrow \quad \alpha_- + \alpha_+ + \alpha_\infty = \pm i (N + 1/2), \quad N = 0, 1, \cdots,
\end{equation}
which leads to
\begin{equation}
\omega = - \frac{B}{Q^2} \alpha_\infty \pm i \left( N + \frac12 \right) \frac{B}{Q^2}, \qquad N = 0, 1, \cdots.
\end{equation}

The extremal limit can be fulfilled by taking limit $B = 0$ and the corresponding KG equations are confluent hypergeometric equations with a degenerate pole at $\rho = 0$. Consequently, the cross section ratio $\sigma_\mathrm{Ex}$ is
\begin{equation} \label{eq:ThyperC}
\sigma_\mathrm{Ex} = \frac{\left| \mathcal{T}_\mathrm{Ex} \right|^2}{\left| \mathcal{I}_\mathrm{Ex} \right|^2} = \mathrm{e}^{-\pi (\alpha_\infty + \tilde{\alpha})} \frac{\sinh(2 \pi \alpha_\infty)}{\cosh(\pi \tilde{\alpha} - \pi \alpha_\infty)},
\end{equation}
and the mean number $\mathcal{N}_\mathrm{Ex}$ is
\begin{equation} \label{eq:NhyperC}
\mathcal{N}_\mathrm{Ex} = \frac{\left| \mathcal{T}_\mathrm{Ex} \right|^2}{\left| \mathcal{R}_\mathrm{Ex} \right|^2} = \mathrm{e}^{\pi (\alpha_\infty - \tilde{\alpha})} \frac{\sinh(2 \pi \alpha_\infty)}{\cosh(\pi \tilde{\alpha} + \pi \alpha_\infty)},
\end{equation}
with $\tilde{\alpha} = q Q$, which again agree with the results in~\cite{Chen:2012zn}.

\subsection{Production of minimal energy particle under static balance in nonextremal black hole} \label{subsec:SpecialCase}
In Ref.~\cite{Hsiao2019}, pair production was studied under a special condition: $\omega = m$ and $m M = q Q$. The condition $m M = q Q$ is the static balance (gravitational and electric forces) of a charge with $m$ and $q$ in the background of nonextremal RN black hole up to the ppN approximation~\cite{Bonnor:1981ag}. The other condition $\omega = m$ is the minimal energy of the particle in the asymptotic region.\footnote{The Hamiltonian for the charge in the metric~(\ref{eq:RNBH}), $H = -m^2 = -\frac{1}{f(r)} (\pi_t - q A_t(r))^2 + f(r) \pi_r^2 + \frac{L^2}{r^2}$, gives the minimal energy when $\pi_t = \omega = m$ at $r = \infty$.} Under the specific condition above, the radial part of the KG equation~(\ref{eq:radialKGeq}) reduces to
\begin{equation} \label{eq:KGNonEx}
\frac{d^2}{d r^2} R + \left( \frac{1}{r - r_+} + \frac{1}{r - r_-} \right) \frac{d}{d r} R + \left[ \left( \frac{\omega r (r_+ - r_-)}{2 (r - r_-) (r - r_+)} \right)^2 - \frac{\lambda_l}{(r - r_-) (r - r_+)} \right] R = 0.
\end{equation}
Certainly, Eq.~\eqref{eq:KGNonEx} belongs to a Riemann differential equation~\eqref{eq:Riemann-equation} and, according to the procedure in the previous subsection, we write down the Riemann P-function as
\begin{equation}
P \begin{pmatrix} r_- & r_+ & \infty & \\ i A_- & i A_+ & 1/2 + i A_\infty & ; r \\ - i A_- & - i A_+ & 1/2 - i A_\infty & \end{pmatrix},
\end{equation}
with
\begin{equation}
2 A_\pm = \omega r_\pm, \qquad A_\infty = \sqrt{\omega^2 (r_+ - r_-)^2/4 - \lambda_l - 1/4}.
\end{equation}
Repeating the same computation in the previous subsection by replacing $\alpha_i$ by $A_i$, the mean number $\mathcal{N}_\mathrm{noF}$ and the absorption cross section ratio $\sigma_\mathrm{noF}$ are
\begin{eqnarray}
&& \mathcal{N}_\mathrm{noF} = \frac{\left| \mathcal{T}_\mathrm{noF} \right|^2}{\left| \mathcal{R}_\mathrm{noF} \right|^2} = \frac{\sinh(2 \pi A_+) \sinh(2 \pi A_\infty)}{\cosh(\pi A_+ + \pi A_- - \pi A_\infty) \cosh(\pi A_+ - \pi A_- - \pi A_\infty)},
\\
&& \sigma_\mathrm{noF} = \frac{\left| \mathcal{T}_\mathrm{noF} \right|^2}{\left| \mathcal{I}_\mathrm{noF} \right|^2} = \frac{\sinh(2 \pi A_+) \sinh(2 \pi A_\infty)}{\cosh(\pi A_+ + \pi A_- + \pi A_\infty) \cosh(\pi A_+ - \pi A_- + \pi A_\infty)}.
\end{eqnarray}
Eqs.~(67) and (68) agree with the results obtained in~\cite{Hsiao2019}, which follows the method in~\cite{Chen:2012zn}. Note that these results apply to pair production of specific (no-force) scalar particles in the full spacetime outside the horizon of RN black holes. This is a new result without any direct comparison with literature. Moreover, this particular case may not be physically interesting due to the strong specific constraints. However, the result may shed light on understanding pair production in non-extremal black holes, which requires a further study.

\section{Pair Production from Generalized P\"oschl-Teller Potential} \label{sec:PoschlTeller}

As the last model with exact wave functions in terms of special functions, we consider the generalized P\"oschl-Teller potential, which was introduced by Rosen and Morse, and apply the monodromy method to find the pair production as well as the absorption cross section in Sec.~\ref{sec:SpecialCases}. This potential occurs for a scalar field in the global de Sitter space and a charged scalar field in a Sauter-type electric field. The wave equation takes the form
\begin{equation} \label{eq_PT}
\psi''(x) + \left( A + B \tanh x + C \tanh^2 x \right) \psi(x) = 0,
\end{equation}
where $\psi$ is a scalar field in the background of curved spaces or electric fields and $x$ denotes a dimensionless space variable.
The equation can be transformed, by using the new coordinate $z \equiv 2/(1 - \tanh x)$, to
\begin{equation}
\psi''(z) + \frac1{z - 1} \psi'(z) + \left( \frac{C}{z^2} + \frac{A - B + C}{4 (z - 1)^2} + \frac{2 C - B}{2 z} - \frac{2 C - B}{2 (z - 1)} \right) \psi(z) = 0,
\end{equation}
which is a Riemann differential equation~\eqref{eq:Riemann-equation} with three simple poles: $z = 0 \, (x = i \pi/2)$, $z = 1 \, (x = -\infty)$, $z = \infty \, (x = \infty)$, and the associated Riemann P-function is
\begin{equation}
P \begin{pmatrix} 0 & 1 & \infty & \\ 1/2 + i \alpha_0 & i \alpha_1 & i \alpha_\infty & ; z \\ 1/2 - i \alpha_0 & -i \alpha_1 & -i \alpha_\infty & \end{pmatrix},
\end{equation}
where
\begin{equation}
\alpha_0 = \sqrt{C - 1/4}, \qquad 2 \alpha_1 = \sqrt{A - B + C}, \qquad 2 \alpha_\infty = \sqrt{A + B + C}.
\end{equation}
Analogous to the previous section, the monodromy matrices are chosen as
\begin{equation}
\begin{split}
& \mathbf{M}_1 = \begin{bmatrix} \mathrm{e}^{-2 \pi \alpha_1} & 1 \\ 0 & \mathrm{e}^{2 \pi \alpha_1} \end{bmatrix}, \qquad \mathbf{M}_\infty = \begin{bmatrix} \mathrm{e}^{-2 \pi \alpha_\infty} & 0 \\ b & \mathrm{e}^{2 \pi \alpha_\infty} \end{bmatrix},
\\
& \mathbf{M}_0 = \begin{bmatrix} \mathrm{e}^{2 \pi (\alpha_1 + \alpha_\infty)} & - \mathrm{e}^{2 \pi \alpha_\infty} \\ -b \, \mathrm{e}^{2 \pi \alpha_1} & b + \mathrm{e}^{-2 \pi (\alpha_1 + \alpha_\infty)} \end{bmatrix},
\end{split}
\end{equation}
with
\begin{equation}
b = - \mathrm{e}^{2 \pi \alpha_0} - \mathrm{e}^{-2 \pi \alpha_0} - \mathrm{e}^{2 \pi (\alpha_1 + \alpha_\infty)} - \mathrm{e}^{-2 \pi (\alpha_1 + \alpha_\infty)}.
\end{equation}
The eigenvectors of $\mathbf{M}_1$ and $\mathbf{M}_\infty$ are
\begin{equation}
\mathbf{E}_1 = \begin{bmatrix} 1 & 1 \\ \mathrm{e}^{2 \pi \alpha_1} - \mathrm{e}^{-2 \pi \alpha_1} & 0 \end{bmatrix}, \qquad \mathbf{E}_\infty = \begin{bmatrix} \mathrm{e}^{-2 \pi \alpha_\infty} - \mathrm{e}^{2 \pi \alpha_\infty} & 0 \\ b & 1 \end{bmatrix},
\end{equation}
from which we get the associated connection matrix
\begin{equation}
\begin{split}
\mathbf{P}_1^\infty &= \begin{bmatrix} d_1 & 0 \\ 0 & d_2 \end{bmatrix} (\mathbf{E}_1)^{-1} \mathbf{E}_\infty \begin{bmatrix} d_3 & 0 \\ 0 & d_4 \end{bmatrix}
\\
&= \frac1{\sinh(2 \pi\alpha_1)} \begin{bmatrix} - d_1 d_3 \left[ \cosh(2 \pi \alpha_0) + \cosh(2 \pi \alpha_1 + 2 \pi \alpha_\infty) \right] & 2 d_1 d_4  \\ d_2 d_3 \left[ \cosh(2 \pi \alpha_0) + \cosh(2 \pi \alpha_1 - 2 \pi \alpha_\infty) \right] & - 2 d_2 d_4 \end{bmatrix}.
\end{split}
\end{equation}
Here, the normalization gives the condition
\begin{equation}
d_1 d_2 d_3 d_4 = \frac{\sinh^2(2 \pi \alpha_0)}{4 \sinh(2 \pi \alpha_1) \sinh(2 \pi \alpha_\infty)}.
\end{equation}
Finally, the absorption cross section can be read out as
\begin{equation} \label{sigmaPT}
\sigma = \frac{\sinh(2 \pi \alpha_1) \sinh(2 \pi \alpha_\infty)}{\cosh(\pi \alpha_0 - \pi \alpha_1 - \pi \alpha_\infty) \cosh(\pi \alpha_0 + \pi \alpha_1 + \pi \alpha_\infty)},
\end{equation}
as well as the mean number
\begin{equation} \label{NPT}
\mathcal{N} = \frac{\sinh(2 \pi \alpha_1) \sinh(2 \pi \alpha_\infty)}{\cosh(\pi \alpha_0 + \pi \alpha_1 - \pi \alpha_\infty) \cosh(\pi \alpha_0 - \pi \alpha_1 + \pi \alpha_\infty)}.
\end{equation}
Actually, from the general formula~\eqref{eq_Nformula} we can easily compute the mean number by substituting $\rho^1 = 1/2 + i \alpha_0, \rho^2 = 1/2 - i \alpha_0, \sigma^1 = i \alpha_1, \sigma^2 = - i \alpha_1$ and $\tau^1 = i \alpha_\infty, \tau^2 = - i \alpha_\infty$.

We now compare the monodromy result with that of the generalized P\"{o}sch-Teller potential in quantum theory. The model explains the Schwinger pair production in a Sauter-type electric field. To do so, we first consider the component equation of the scalar field in a background field~\cite{Bagrov:1990}
\begin{equation} \label{eq_tPT}
\psi''(x) + k^2(x) \psi(x) = 0,
\end{equation}
where the momentum square is
\begin{equation}
k^2(x) = k_-^2 + \frac12 (k_+^2 - k_-^2) \left( 1 + \tanh\frac{x}{L} \right) + \frac{V}{4} \left( \tanh^2\frac{x}{L} - 1 \right).
\end{equation}
Here, $L$ is a characteristic length scale that makes $x/L$ dimensionless, and is an effective length scale for the Sauter-type electric field $E(x) = E_0/\cosh^2(x/L)$.

The momentum has two limits in the far left and right: $k(-\infty) = k_-$ and $k(+\infty) = k_+$. The solution that has a unit incident flux $x = - \infty$ and only out-going flux at $x = \infty$ is given by
\begin{equation}
\psi(x) \!=\! \left( 1 \!+\! \tanh\frac{x}{L} \right)^{-i k_- L/2} \left( 1 \!-\! \tanh\frac{x}{L} \right)^{-i k_+ L/2} F\left( \frac{\mu}2, \frac{\nu}2, 1 \!-\! i k_+ L; \frac{1 \!-\! \tanh\frac{x}{L}}{2} \right),
\end{equation}
where
\begin{equation}
\mu = 1 - i L (k_- + k_+) - i \Delta, \qquad \nu = 1 - i L (k_- + k_+) + i \Delta.
\end{equation}
Here, $\Delta = \sqrt{V L^2 - 1}$.
Using the transformation formula 9.131.2 of~\cite{Gradshteyn:2014}, we find the asymptotic forms of the wave function, with normalization $I = 1$, \begin{equation} \label{space-scattering}
\psi(x \rightarrow -\infty) = \mathrm{e}^{ i k_- x} + R \, \mathrm{e}^{- i k_- x}, \qquad \psi(x \rightarrow \infty) = T \, \mathrm{e}^{ i k_+ x},
\end{equation}
where
\begin{eqnarray}
R &=& \frac{\Gamma\bigl( i k_- L \bigr) \, \Gamma\Bigl( \frac12 - \frac{i}2 (k_+ L + k_- L + \Delta) \Bigr) \, \Gamma\Bigl( \frac12 - \frac{i}2 (k_+ L + k_- L - \Delta) \Bigr)}{\Gamma\bigl( -i k_- L \bigr) \, \Gamma\Bigl( \frac12 - \frac{i}2 (k_+ L - k_- L + \Delta) \Bigr) \, \Gamma\Bigl( \frac12 - \frac{i}2 (k_+ L - k_- L + \Delta) \Bigr)},
\\
T &=& \frac{\Gamma\Bigl( \frac12 - \frac{i}2 (k_+ L + k_- L + \Delta) \Bigr) \, \Gamma\Bigl( \frac12 - \frac{i}2 (k_+ L + k_- L - \Delta) \Bigr)}{\Gamma\bigl( 1 - i k_+ L \bigr) \, \Gamma\bigl( -i k_- L \bigr)},
\end{eqnarray}
The Bogoliubov coefficients are related to fluxes~\cite{Chen:2012zn} as $| \alpha |^2 = \mathrm{(incident~flux)}/\mathrm{(reflection~flux)} = 1/| R |^2$ and $| \beta |^2 = \mathrm{(transmission~flux)}/\mathrm{(reflection~flux)} = | T |^2/| R |^2$ which lead to
\begin{eqnarray}
| \alpha |^2 = \frac{\cosh\left[ \frac{\pi}{2} (k_+ L + k_- L + \Delta) \right] \cosh\left[ \frac{\pi}{2} (k_+ L + k_- L - \Delta) \right]}{\cosh\left[ \frac{\pi}{2} (k_+ L - k_- L + \Delta) \right] \cosh\left[ \frac{\pi}{2} (k_+ L - k_- L - \Delta) \right]}, \label{alphaRM}
\\
| \beta |^2 = \frac{\sinh(\pi k_- L) \sinh(\pi k_+ L)}{\cosh\left[ \frac{\pi}{2} (k_+ L - k_- L + \Delta) \right] \cosh\left[ \frac{\pi}{2} (k_+ L - k_- L - \Delta) \right]}. \label{betaRM}
\end{eqnarray}
The Bogoliubov coefficients satisfy the relation $| \alpha |^2 - | \beta |^2 = 1$ for bosons. The absorption cross section defined as $\sigma = (\mathrm{tranmission~flux})/(\mathrm{incident~flux}) = | T |^2 = | \beta |^2/| \alpha |^2$ is
\begin{eqnarray} \label{saut-abs}
\sigma = \frac{\sinh(\pi k_- L) \sinh(\pi k_+ L)}{\cosh\left[ \frac{\pi}{2} (k_+ L + k_- L + \Delta) \right] \cosh\left[ \frac{\pi}{2} (k_+ L + k_- L - \Delta) \right]}.
\end{eqnarray}

By comparing equation~(\ref{eq_tPT}) with the equations~(\ref{eq_PT}), we can identify
\begin{equation}
A = L^2 [ (k_+^2 + k_-^2)/2 - V/4 ], \qquad B = - L^2 (k_+^2 - k_-^2)/2, \qquad C = V L^2/4,
\end{equation}
and thereby the corresponding parameters
\begin{equation}
\begin{split}
& \alpha_0 = \sqrt{C - 1/4} = \frac12 \sqrt{V L^2 - 1} = \Delta/2, \qquad 2 \alpha_1 = \sqrt{A - B + C} = k_+ L,
\\
& 2 \alpha_\infty = \sqrt{A + B + C} = k_- L.
\end{split}
\end{equation}
Remarkably, substituting these parameters into Eqs.~(\ref{sigmaPT}) and (\ref{NPT}), the result from the monodromy is identical to Eqs.~(\ref{saut-abs}) and (\ref{betaRM}) and consistent with the result of~\cite{Chervyakov:2009bq, Kim:2009pg} for scalar pair production in the Sauter-type electric field with $V/4 = (q E_0 L)^2$. Note that for spin-1/2 fermions, setting $V/4 = (q E_0 L)^2 \pm i q E_0$ in~(\ref{sigmaPT}) gives the correct result for the mean number of fermions  in the Sauter-type electric field~\cite{Kim:2009pg, Nikishov:1970br, Hansen:1980nc}. A passing remark is the phase-integral method that also makes use of the regular singular points gives the leading  Boltzmann factor for the pair production in the Sauter-type electric field~\cite{Kim:2007pm}. We argue that the monodromy may complete the phase-integral method.

\section{Conclusion} \label{sec:Conclusion}
In this paper we have advanced the monodromy approach to find the exact mean number of the pair production in background fields whose mode equations belong to the Riemann differential equation. Further, we have investigated the pair production of charged scalar in two special cases of RN black holes and in the generalized P\"oschl-Teller potential. In these three cases the mode equations after separating conserved quantum numbers are (confluent) hypergeometric functions, the specific case of the Riemann differential equation, from which scattering coefficients and therefrom the associated Bogolubov coefficients have been directly computed. The monodromy technique provides a practical and powerful method to compute the mean number and absorption cross section while it avoids cumbersome and subtle transformation formulas of the hypergeometric functions that connect different regions related to the singular points of the underlying equation. Instead, we only need to compute the exponents of the Frobenius solutions at relevant singular points, which significantly simplifies computation.

As an important future work, we will study the Schwinger pair production and Hawking radiation in the general RN black holes, in which the relevant mode equation is a confluent Heun equation with an irregular singularity at infinity. Resolving the irregular singularity and regular poles more than three is a challenging task in applying the monodromy approach. A naive idea is firstly to apply the monodromy technique for the Heun equation with four simple poles, and then take a proper limit to get the results for confluent Heun equations, analogously from Eqs.~(\ref{eq:Thyper}),~(\ref{eq:Nhyper}) to Eqs.~(\ref{eq:ThyperC}),~(\ref{eq:NhyperC}). Of course, we have to deal with some new subtleties, such as appearance of accessary parameters and derivation of monodromy matrices etc. Hopefully the results of the no-force case in Sec.~\ref{subsec:SpecialCase} may provide in part some clues for resolving possible ambiguities. Nevertheless, even without closed forms, developing rigorous numerics will contribute to physical problems under study.

\acknowledgments
C.M.C. would like to appreciate the hospitality at Kunsan National University and CQUeST, Sogang University, where part of the revision was made.
The work of C.M.C. was supported by the National Science and Technology Council of the R.O.C. (Taiwan) under the grants MOST 111-2112-M-008-012 and 110-2923-M-002-016-MY3.
The work of S.P.K. was supported in part by National Research Foundation of Korea (NRF) funded by the Ministry of Education (2019R1I1A3A01063183).
AT was supported by JSPS KAKENHI Grant Numbers JP22K03411 and JP21H01001.

\begin{appendix}

\end{appendix}


\begin{thebibliography}{99}

\bibitem{Schwinger:1951nm}
J.~S.~Schwinger,
``On gauge invariance and vacuum polarization,''
Phys. Rev. \textbf{82} (1951), 664-679.
doi:10.1103/PhysRev.82.664

\bibitem{Sauter:1931zz}
F.~Sauter,
``Uber das Verhalten eines Elektrons im homogenen elektrischen Feld nach der relativistischen Theorie Diracs,''
Z. Phys. \textbf{69} (1931), 742-764.
doi:10.1007/BF01339461

\bibitem{Hawking:1975vcx}
S.~W.~Hawking,
``Particle Creation by Black Holes,''
Commun. Math. Phys. \textbf{43} (1975), 199-220
[erratum: Commun. Math. Phys. \textbf{46} (1976), 206]
doi:10.1007/BF02345020

\bibitem{DeWitt:2003}
B.~S.~DeWitt,
{\sl The global approach to quantum field theory},
(Oxford University Press, USA, 2003).

\bibitem{Bagrov:1990}
V.~G.~Bagrov and D.~Gitman,
{\sl Exact solutions of relativistic wave equations},
(Springer Science \& Business Media, 1990).

\bibitem{Bender:1999}
C.~M.~Bender, S.~Orszag and S.~A.~Orszag,
{\sl Advanced mathematical methods for scientists and engineers I: Asymptotic methods and perturbation theory},
(Springer Science \& Business Media, 1999).

\bibitem{Honda:2015}
N.~Honda, T.~Kawai and Y.~Takei,
{\sl Virtual turning points},
(Springer, Tokyo, 2015).

\bibitem{Parker:1968mv}
L.~Parker,
``Particle creation in expanding universes,''
Phys. Rev. Lett. \textbf{21} (1968), 562-564.
doi:10.1103/PhysRevLett.21.562

\bibitem{Ford:2021syk}
L.~H.~Ford,
``Cosmological particle production: a review,''
Rept. Prog. Phys. \textbf{84} (2021) no.11, 116901
doi:10.1088/1361-6633/ac1b23
[arXiv:2112.02444 [gr-qc]].

\bibitem{Ruffini:2009hg}
R.~Ruffini, G.~Vereshchagin and S.~S.~Xue,
``Electron-positron pairs in physics and astrophysics: from heavy nuclei to black holes,''
Phys. Rept. \textbf{487} (2010), 1-140
doi:10.1016/j.physrep.2009.10.004
[arXiv:0910.0974 [astro-ph.HE]].

\bibitem{Fedotov:2022ely}
A.~Fedotov, A.~Ilderton, F.~Karbstein, B.~King, D.~Seipt, H.~Taya and G.~Torgrimsson,
``Advances in QED with intense background fields,''
Phys. Rept. \textbf{1010}, 1-138 (2023)
doi:10.1016/j.physrep.2023.01.003
[arXiv:2203.00019 [hep-ph]].

\bibitem{Chen:2012zn}
C.-M.~Chen, S.~P.~Kim, I.-C.~Lin, J.-R.~Sun and M.-F.~Wu,
``Spontaneous Pair Production in Reissner-Nordstrom Black Holes,''
Phys. Rev. D \textbf{85} (2012), 124041
doi:10.1103/PhysRevD.85.124041
[arXiv:1202.3224 [hep-th]].

\bibitem{Chen:2016caa}
C.-M.~Chen, S.~P.~Kim, J.-R.~Sun and F.-Y.~Tang,
``Pair Production in Near Extremal Kerr-Newman Black Holes,''
Phys. Rev. D \textbf{95} (2017) no.4, 044043
doi:10.1103/PhysRevD.95.044043
[arXiv:1607.02610 [hep-th]].

\bibitem{Chen:2017mnm}
C.-M.~Chen, S.~P.~Kim, J.-R.~Sun and F.-Y.~Tang,
``Pair production of scalar dyons in Kerr\textendash{}Newman black holes,''
Phys. Lett. B \textbf{781} (2018), 129-138
doi:10.1016/j.physletb.2018.03.078
[arXiv:1705.10629 [hep-th]].

\bibitem{Chen:2020mqs}
C.-M.~Chen and S.~P.~Kim,
``Schwinger Effect from Near-extremal Black Holes in (A)dS Space,''
Phys. Rev. D \textbf{101} (2020) no.8, 085014
doi:10.1103/PhysRevD.101.085014
[arXiv:2002.00394 [hep-th]].

\bibitem{Zhang:2020apg}
J.~Zhang, Y.-Y.~Lin, H.-C.~Liang, K.-J.~Chi, C.-M.~Chen, S.~P.~Kim and J.-R.~Sun,
``Pair production in Reissner-Nordstr\"om-Anti de Sitter black holes,''
Chin. Phys. C \textbf{45} (2021) no.6, 065105
doi:10.1088/1674-1137/abf4f6
[arXiv:2003.06398 [hep-th]].

\bibitem{Cai:2020trh}
R.-G.~Cai, C.-M.~Chen, S.~P.~Kim and J.-R.~Sun,
``Schwinger effect in near-extremal charged black holes in high dimensions,''
Phys. Rev. D \textbf{101} (2020) no.10, 105015
doi:10.1103/PhysRevD.101.105015
[arXiv:2004.00735 [hep-th]].

\bibitem{Chen:2021jwy}
C.-M.~Chen and S.~P.~Kim,
``Dyon production from near-extremal Kerr\textendash{}Newman\textendash{}(anti-)de Sitter black holes,''
Eur. Phys. J. C \textbf{83} (2023) no.3, 219
doi:10.1140/epjc/s10052-023-11380-7
[arXiv:2111.14124 [hep-th]].

\bibitem{Hsiao2019}
B.~Hsiao,
``Pair Production in Charged Black Holes Admitting Heun-to-Hypergeometric Reduction,''
MSc. Thesis, National Central University, 2019

\bibitem{IKSY2013}
K.~Iwasaki, H.~Kimura, S.~Shimomura and M.~Yoshida,
{\sl From Gauss to Painlev\'e: a modern theory of special functions},
(Braunschweig: Vieweg, 2012).

\bibitem{Castro:2013lba}
A.~Castro, J.~M.~Lapan, A.~Maloney and M.~J.~Rodriguez,
``Black Hole Scattering from Monodromy,''
Class. Quant. Grav. \textbf{30} (2013), 165005
doi:10.1088/0264-9381/30/16/165005
[arXiv:1304.3781 [hep-th]].


\bibitem{Kim:2007pm}
S.~P.~Kim and D.~N.~Page,
``Improved Approximations for Fermion Pair Production in Inhomogeneous Electric Fields,''
Phys. Rev. D \textbf{75} (2007), 045013
doi:10.1103/PhysRevD.75.045013
[arXiv:hep-th/0701047 [hep-th]].

\bibitem{Kim:2013cka}
S.~P.~Kim,
``Geometric Origin of Stokes Phenomenon for de Sitter Radiation,''
Phys. Rev. D \textbf{88}, no.4, 044027 (2013)
doi:10.1103/PhysRevD.88.044027
[arXiv:1307.0590 [hep-th]].

\bibitem{Hansen:1980nc}
A.~Hansen and F.~Ravndal,
``Klein's Paradox and Its Resolution,''
Phys. Scripta \textbf{23}, 1036 (1981)
doi:10.1088/0031-8949/23/6/002

\bibitem{Kim:2000un}
S.~P.~Kim and D.~N.~Page,
``Schwinger pair production via instantons in a strong electric field,''
Phys. Rev. D \textbf{65}, 105002 (2002)
doi:10.1103/PhysRevD.65.105002
[arXiv:hep-th/0005078 [hep-th]].

\bibitem{Brout:1995rd}
R.~Brout, S.~Massar, R.~Parentani and P.~Spindel,
``A Primer for black hole quantum physics,''
Phys. Rept. \textbf{260}, 329-454 (1995)
doi:10.1016/0370-1573(95)00008-5
[arXiv:0710.4345 [gr-qc]].

\bibitem{Bonnor:1981ag}
W.~B.~Bonnor,
``The Equilibrium of Two Charged Masses in General Relativity,''
Phys. Lett. A \textbf{83} (1981), 414-416
doi:10.1016/0375-9601(81)90467-9

\bibitem{Gradshteyn:2014}
I.~S.~Gradshteyn and I.~M.~Ryzhik,
{\sl Table of integrals, series, and products},
(Academic press, 2014)

\bibitem{Chervyakov:2009bq}
A.~Chervyakov and H.~Kleinert,
``Exact Pair Production Rate for a Smooth Potential Step,''
Phys. Rev. D \textbf{80} (2009), 065010
doi:10.1103/PhysRevD.80.065010
[arXiv:0906.1422 [hep-th]].

\bibitem{Kim:2009pg}
S.~P.~Kim, H.~K.~Lee and Y.~Yoon,
``Effective Action of QED in Electric Field Backgrounds II. Spatially Localized Fields,''
Phys. Rev. D \textbf{82} (2010), 025015
doi:10.1103/PhysRevD.82.025015
[arXiv:0910.3363 [hep-th]].

\bibitem{Nikishov:1970br}
A.~I.~Nikishov,
``Barrier scattering in field theory removal of klein paradox,''
Nucl. Phys. B \textbf{21} (1970), 346-358.
doi:10.1016/0550-3213(70)90527-4

\end{thebibliography}
\end{document}